%% file: main.tex
\documentclass[conference,compsoc]{IEEEtran}
\ifCLASSOPTIONcompsoc
  \usepackage[nocompress]{cite}
\else
  \usepackage{cite}
\fi
\usepackage[english]{babel} 
\usepackage{tikz}
\usepackage{xurl}
\usepackage{amsmath}
\usepackage{mathtools}
\usepackage{pifont}
\usepackage{algorithm}
\usepackage{algorithmicx}
\usepackage{algpseudocode}
\usepackage{amsmath}
\usepackage{amsthm}
\usepackage{enumitem}
\usepackage{microtype}
\usepackage{graphicx}
\usepackage{wasysym}
\usepackage{soul}

\usepackage{bbding}
\usepackage{subfigure}
\usepackage{booktabs} 
\usepackage[table]{xcolor}
 \usepackage{balance}
 \usepackage{multirow}
\usepackage{multicol}
\usepackage{makecell}
\usepackage{hyperref}
\usepackage[all]{nowidow}
\usepackage{color, colortbl} 
\usepackage{xspace}
\usepackage{multirow}
\usepackage{tcolorbox}
\usepackage{csquotes}

\usepackage{array}

\newcommand{\outmark}[1]{\makebox[1.25ex][c]{#1}}
\newcommand{\greensq}{\outmark{$\circ$}}
\newcommand{\redsq}{\outmark{$\bullet$}}
\newcommand{\graysq}{\outmark{--}}

\definecolor{Gray}{gray}{0.9}

\definecolor{payloadcolor}{HTML}{FCEBEB}
\definecolor{cleancolor}{HTML}{EAF3DE}
\definecolor{dropcolor}{HTML}{F1EFE8}

\usepackage[available,functional,reproduced]{ieeebadges}

\begin{document}

%
 \title{What Users See Is Not What Models Read: Split-View PDFs in \\ Document-to-LLM Supply Chains}

\author{\IEEEauthorblockN{Side Liu}
\IEEEauthorblockA{Tulane University}
\and
\IEEEauthorblockN{Jiang Ming}
\IEEEauthorblockA{Tulane University}}

\maketitle

\pagestyle{plain}
\thispagestyle{plain}


\input{pages/abstract}


%
\IEEEpeerreviewmaketitle

\input{pages/introduction}

\input{pages/background}
\input{pages/threatmodel}

\input{pages/taxonomy}

\input{pages/evaluation}
\input{pages/discussion}






%




\section*{Acknowledgments}
We thank the anonymous IEEE S\&P 2027 reviewers for their valuable comments that improved this paper.
This work was supported by NSF grants 2417055 and 2555093 and a Google Research Scholar Award.

\bibliographystyle{IEEEtran}
\bibliography{refs/fuzzing,refs/online,refs/pdfs}

\input{pages/ethics}
\input{pages/appendix}

\end{document}

%% file: pages/abstract.tex
\begin{abstract}

Document-to-LLM applications typically read uploaded PDFs by first
translating them into text through an extraction layer whose output is
not ordinarily exposed to users.
We show that this layer enables \textit{split-view PDFs}: one document can have two semantic views before model reasoning. By mining specification-permitted or implementation-tolerated representation gaps at the PDF render/extract boundary, we instantiate 25 extraction gaps (EG) in which extractors return attacker-controlled or extractor-dependent text while the rendered page shows benign or different content. The gaps form four families: semantic overrides, hidden semantic injection, reading-order splits, and font-decoding splits, and 14 gaps have no exact path/mechanism-level match in prior PDF-to-LLM attacks.

We evaluate these gaps on 16 PDF processing stacks and 7 commercial LLM services that accept PDFs through official APIs and web chat. Each gap causes render-extract divergence on at least one stack. Under a gap-level exposure criterion, every evaluated service exposes at least one gap, with 12/25 to 21/25 exposed gaps. Across deployment variants, exposure is driven mainly by the ingestion stack---not model identity alone---that constructs the model’s document context: APIs, cloud backends, web frontends, and local runtimes can expose different views of the same PDF. We further show that tested safety filters cover only selected hidden-text constructions. To support triage, we also develop a static screening scanner whose rules trigger on all 25 benchmark gaps in our self-test, and we discuss dual-view consistency as a longer-term defense direction.

\end{abstract}

%% file: pages/introduction.tex
\section{Introduction}
\label{sec:introduction}

The Portable Document Format (PDF) dominates document exchange
worldwide~\cite{pdfa_popularity,lam2024pdf}.
LLM applications now routinely accept PDF uploads for summarization,
question answering, structured extraction, and review~\cite{xie2024wukong,AIChatwithPDF,ChatPDF,devNvidiaPDF,nvida_rag,LlamaIndex}. 
Across official APIs, web chat, and document-analysis products, they
share a document-ingestion layer that converts PDFs into model-consumed
text before reasoning. Users do not ordinarily see the resulting
extractor-side text alongside the rendered document.
Document-to-LLM pipelines therefore
assume that the extracted text consumed by the LLM reflects the
meaning visible to the user.

\begin{figure}[]
    \centering
    \includegraphics[width=0.95\linewidth]{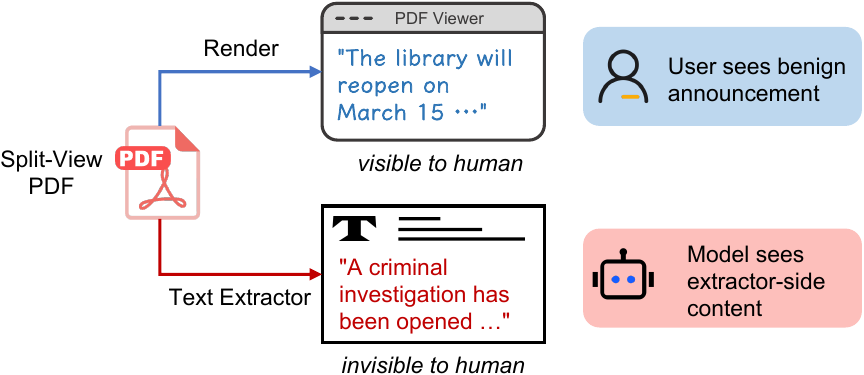}
   \vspace{-2mm}
    \caption{A split-view PDF creates a mismatch between the user-visible page and the model-consumed representation.}
    \label{fig:introduction}
    \vspace{-4mm}
\end{figure}

That assumption is fragile because PDF separates visual rendering from
text extraction, and extractors differ in how they resolve ambiguous
or optional structures~\cite{lam2024pdf,blecher2023nougat,
siebenschuh2025adaparse}. Figure~\ref{fig:introduction} illustrates
the core primitive: a \emph{split-view PDF} whose rendered page appears
benign while the extracted text carries attacker-controlled or
extractor-dependent semantics. This is a document-ingestion
supply-chain failure upstream of model reasoning: the hidden ingestion
layer constructs a document context the user cannot observe.
Prior work~\cite{greshake2023not,
li2026making,jin2025trapdoc,xiong2025invisible,luo2026pdf} has exposed related risks: untrusted external content can
steer LLM-integrated applications, visual and font-based encodings can
create human-machine perception gaps, and PDF-specific hidden-text or obfuscation techniques can reach the model through document-ingestion pipelines. 
Together, these works show that adversarial document content can reach LLM workflows, but leave the ingestion layer largely as a black box: they do not characterize the PDF render/extract mechanisms, parser fallbacks, and deployment choices that construct a model-consumed view different from the user-visible page.

In this paper, we mine specification-permitted or implementation-tolerated render/extract
divergence points in ISO~32000 rather than enumerate
isolated tricks. Starting from places where semantic layers, visibility
state, reading order, or font metadata allow renderers and extractors
to diverge, we instantiate candidates as canaries and keep only those
with stable rendering and attacker-controllable extractor divergence.
This process yields 25 confirmed extraction
gaps (EG) across four families---semantic overrides, hidden semantic
injection, reading-order splits, and font-decoding splits---including
14 gaps with no direct counterpart in prior PDF-to-LLM attack work.
Several gaps require no hidden text: the page is fully visible, yet
extraction returns different semantics. Hidden-text filters do not
address such cases because there is no hidden payload text to remove.
We organize the gap set into a two-tier benchmark: mechanism canaries for parser and
ingestion-stack measurement, and end-to-end attack documents for LLM
services and applications. We evaluate the benchmark across 16 PDF
processing stacks and 7 commercial LLM services, then compare official
APIs, cloud-hosted backends, web chat products, and local runtimes for
the same nominal models. The results show that every gap produces
render-extract divergence on at least one stack. Our evaluation results
show that every evaluated LLM service is affected, with EG-level
coverage ranging from 12/25 to 21/25 gaps and many successful cells
repeating across trials. Exposure is primarily shaped
by the ingestion stack that constructs the model's document context. The
same canaries also provide a side channel for fingerprinting hidden
ingestion behavior: their parser-dependent outputs reveal which PDF
parser or loader-equivalent path is likely used by an LLM application
or RAG pipeline within our controlled panel.

In a nutshell, we make the following key contributions:

\begin{itemize}

\item We systematically identify and categorize 25 PDF extraction gaps across
four root-cause families, including 14 with no direct counterpart
in prior studies. 

\item We build a two-tier benchmark for testing PDF-ingestion
semantic integrity in LLM services and applications. Mechanism
canaries isolate each extraction gap at the parser layer, while
end-to-end attack documents test whether the gap propagates into
model outputs.

\item We conduct a large-scale measurement across 16 PDF processing
stacks and 7 commercial LLMs, comparing official APIs,
cloud-hosted backends, web chat products, and local runtimes. The
results show that PDF ingestion is a supply-chain security boundary:
exposure is shaped by parser, converter, and deployment choices, not
by the model name alone.

\item We show that benchmark canaries support
fingerprinting of PDF ingestion stacks, allowing auditors to infer the
parser or loader-equivalent path used by LLM applications within our controlled panel. We also evaluate current PDF
safety filters and develop a static scanner.

\end{itemize}

%% file: pages/background.tex
\section{Background \& Related Work}
\label{sec:background}

\subsection{The PDF Dual-Path Problem}
\label{sec:bg:pdf}

The PDF format, standardized as ISO~32000~\cite{pdf_standard}, separates the concerns of visual rendering and text extraction into two independent processing paths. A PDF page's content stream contains drawing operators that position and render characters using specific fonts. The visual appearance of each character is determined by the font's glyph program or glyph outlines. Text extraction, the process of recovering Unicode text from a PDF, follows an entirely separate path.

\vspace*{2pt}
\noindent\textbf{Text Extraction Resolution Hierarchy }
When a text extractor encounters a character code in a content stream, it must resolve the corresponding Unicode character. The PDF specification defines a resolution hierarchy: (1) if a \texttt{/ToUnicode} CMap is present, use it to map the character code directly to a Unicode string; (2) if \texttt{/ToUnicode} is absent, fall back to the font's \texttt{/Encoding} dictionary and \texttt{/Differences} array to obtain glyph names, then infer Unicode from the Adobe Glyph List; (3) if no encoding information is available, use the raw character code as a Unicode code point. Each fallback level introduces extractor-dependent behavior, since different implementations resolve ambiguities differently.

\vspace*{2pt}
\noindent\textbf{The Independence Property }
Critically, \texttt{/ToUnicode} mapping is entirely independent of the font's visual rendering. A \texttt{/ToUnicode} CMap can map a character code to any arbitrary Unicode string regardless of what glyph the font program draws for that code. This independence is by design: the PDF specification introduced \texttt{/ToUnicode} to support search and accessibility for fonts with non-standard encodings~\cite{pdf_standard}. However, it means that a font can be constructed where every glyph visually displays one character while the \texttt{/ToUnicode} mapping returns a completely different string.

\vspace*{2pt}
\noindent\textbf{Semantic Markup Path }
Beyond font-level mappings, PDF provides document-level mechanisms that can override text extraction. The \texttt{/ActualText} attribute on marked-content sequences specifies the text that a content span represents~\cite{pdf_standard}. Extractors that implement accessibility-aware text extraction substitute the \texttt{/ActualText} value for the content-stream character codes. Since \texttt{/ActualText} is invisible in the rendered view, existing only in the PDF's logical structure tree, it creates a channel for divergence between visible and extracted content.

\vspace*{2pt}
\noindent\textbf{Format Design Versus Pipeline Assumption }
The independence of rendering and extraction paths is not a defect of
the PDF specification. PDF predates LLM ingestion, and its parallel
paths support legitimate accessibility, search, and copy-paste use
cases. The semantic integrity gap we study is not that rendering or
extraction operations are individually broken, but that ISO~32000 does
not require their outputs to remain mutually consistent at a
document-to-LLM pipeline boundary. That boundary did not exist when the
standard was written, and deployed document-to-LLM systems still do not
enforce it today.

\subsection{PDF Text Extractors in LLM Pipelines}
\label{sec:bg:extractors}

Modern LLM applications employ a variety of PDF tools, which we categorize into three classes.

\vspace*{2pt}
\noindent\textbf{Rule-Based Extractors }
Libraries such as pypdf~\cite{pypdf}, pdfminer~\cite{pdfminer},
MuPDF~\cite{mupdf}, Poppler~\cite{poppler},
PDFium~\cite{pdfium}, PDF.js~\cite{pdfjs}, and Xpdf~\cite{xpdf} parse PDF object
structures and follow the extraction path described above. Each library
independently implements the resolution hierarchy, with different fallback
behaviors for ambiguous or absent fields. These implementation differences
are the primary source of extraction gaps exploited in this work.

\vspace*{2pt}
\noindent\textbf{Rendered-Page Extraction }
Some document pipelines first render PDF pages to images and recover
content from the rendered-page representation through OCR, vision models,
or hybrid converters such as Docling~\cite{docling2025} and Unstructured~\cite{unstructured}.
This approach is designed to resist text-level
split-view attacks because the model or OCR system observes the same
pixel representation as the human. 
However, as we observe in one
commercial web frontend (\S\ref{sec:eval:defense}), this resistance can be 
conditional: extractor-side content is absent for shorter PDFs but reappears at longer pages, 
the platform may switch to text extraction, re-exposing the attack surface it avoids.

\vspace*{2pt}
\noindent\textbf{AI-Oriented Extractors }
Tools designed specifically for LLM pipelines, including
LlamaParse~\cite{LlamaIndex} and OpenDataLoader~\cite{opendataloader},
often wrap rule-based extractors with additional preprocessing.
OpenDataLoader advertises built-in rendering-mismatch safety filters for
PDF prompt-injection content. We evaluate the tested default
configuration as part of our defense analysis in \S\ref{sec:eval:defense}.

\subsection{Adjacent Threats and Scope}
\label{sec:bg:related}

\noindent\textbf{Indirect Prompt Injection and Document Threats }
Indirect prompt-injection work shows that LLM-integrated systems can be
steered by attacker-controlled external content such as web pages,
emails, or retrieved documents~\cite{greshake2023not}. Follow-up
document attacks instantiate this threat in PDF and RAG settings using
hidden prompts, low-visibility text, metadata, Unicode tricks, and
loader-specific behaviors~\cite{jin2025trapdoc,castagnaro2025hidden,
collu2025misleading,pdfpromptinjectiontoolkit}. Together, they show
that attacker-controlled content embedded in documents can enter LLM
workflows and alter model outputs.
We study how PDF documents create that divergent input path: PDF structures and tolerated fallback paths can supply model-consumed
text that differs from the page a
user inspected.

\vspace*{2pt}
\noindent\textbf{PDF Visibility and Representation Attacks }
PDF-specific work has long exploited differences between visual content
and machine-consumed content. PDF Mirage masks content for online
services by manipulating font/glyph interpretation~\cite{markwood2017pdfmirage}.
Recent PDF-to-LLM work extends this direction with obfuscation,
absolute-position reordering, hidden text, and font-data
manipulation~\cite{luo2026pdf,jin2025trapdoc,xiong2025invisible}. Related visual-smuggling work shows that human and model views can diverge even outside the PDF parser stack~\cite{li2026making}. 
These works demonstrate that individual PDF carriers can hide, reorder, or reshape content before it reaches a model. We study a different unit of failure: extraction gaps, i.e., reproducible render/extract divergences that make the ingestion stack construct a model-consumed document different from the user-visible one. This shift separates hidden-text delivery from broader semantic-integrity failures where no hidden payload text needs to be present.

\vspace*{2pt}
\noindent\textbf{Other PDF Security Research }
PDF security research also spans several adjacent but distinct problems:
content-integrity failures in signed documents, including approval-signature
bypasses, shadow attacks, and certification-signature attacks
~\cite{pdfsignatureccs2019,shadowattacks2021,pdfcertification2021}; PDF
malware analysis and detection
~\cite{carmony2016extract,xu2016automatically,chen2020training,VAPD,PDFObj2Vec};
and PDF parser differential and compatibility studies
~\cite{petsios2017nezha,lam2024pdf}.
These works target signature validation, malware
classification, or parser robustness. Our scope is different:
document-to-LLM semantic integrity, where the question is whether the text
constructed by a PDF ingestion stack matches the page a user inspected.


%% file: pages/threatmodel.tex
\section{Threat Model}
\label{sec:threat}

\noindent\textbf{Attacker }
An attacker controls a PDF document that enters a
document-to-LLM pipeline through any normal ingestion path. The attacker
has no access to model weights, system prompts, internal
implementations, or preprocessing stacks. Scenarios
include RAG systems that index external documents, user-uploaded
third-party PDFs for document review, and enterprise workflows ingesting
external invoices, contracts, compliance submissions, or review
manuscripts.
The attacker does \emph{not} control the user's task prompt, such as
``summarize this document.'' The attacker controls only the PDF file and
its objects, which can cause the model-facing PDF preprocessing stack to
output text that differs from the rendered page the user inspected. In
our evaluation, the attacker-controlled strings are benign claims or
answer spans rather than jailbreaks or imperative instructions.

The attacker's primary goal is to induce a downstream
\emph{semantic-integrity violation}: model output becomes unfaithful to
the human-visible document by incorporating attacker-controlled claims,
contradicting or omitting visible claims, or acting on
extractor-dependent artifacts. The same extraction gaps could also carry
malicious instructions and serve as carriers for indirect prompt
injection, but whether a model follows, refuses, or is jailbroken by such
instructions is a downstream robustness question outside this work.
Visible-text prompt injection~\cite{liu2024formalizing}, LLM
jailbreaks~\cite{wei2023jailbroken}, JavaScript or Actions~\cite{vsrndic2013detection}, and memory exploitation~\cite{szekeres2013sok} are also out of scope. The same
canaries support \emph{ingestion-stack fingerprinting}
(\S\ref{sec:eval:fingerprint}) by exposing hidden extraction behavior.

\vspace*{2pt}
\noindent\textbf{Security Property }
We formalize the underlying \emph{semantic integrity} assumption. Let
$V(D)$ be the human-visible claim set and $X_e(D)$ the machine-consumed
claim set produced by extractor $e$. \textit{Soundness} requires
$X_e(D)$ to contain no material claim absent from or contradicting
$V(D)$ (\emph{claim injection}, \emph{claim contradiction}).
\textit{Completeness} requires $X_e(D)$ not to omit any element of
$V(D)$ whose absence would change a downstream task outcome
(\emph{claim omission}). The same properties apply at the model-output
level, with pipeline output $M_p(D,q)$ under task prompt $q$ in place of
$X_e(D)$.
We operationalize materiality at the task level: a claim is material if
its presence, absence, or contradiction changes the labeled summary fact
set or QA answer.

%% file: pages/taxonomy.tex
\section[Split-View PDFs: Construction and Benchmark]{Split-View PDFs: Construction and\\ Benchmark}
\label{sec:taxonomy}

\subsection{Root Causes and Mechanism Families}
\label{sec:families}

Split-view PDFs arise not from implementation bugs in any single tool,
but from specification-permitted representation or implementation-tolerated gaps in PDF's dual
rendering and extraction model. ISO~32000 permits renderers and
extractors to consume different document representations without
requiring their outputs to remain mutually consistent. We group the
confirmed gaps into four root-cause families.
We use \emph{EG} as shorthand for extraction gap: EG01--EG25 are stable catalog identifiers, not severity rankings.

\begin{figure}[]
    \centering
    \includegraphics[width=1\linewidth]{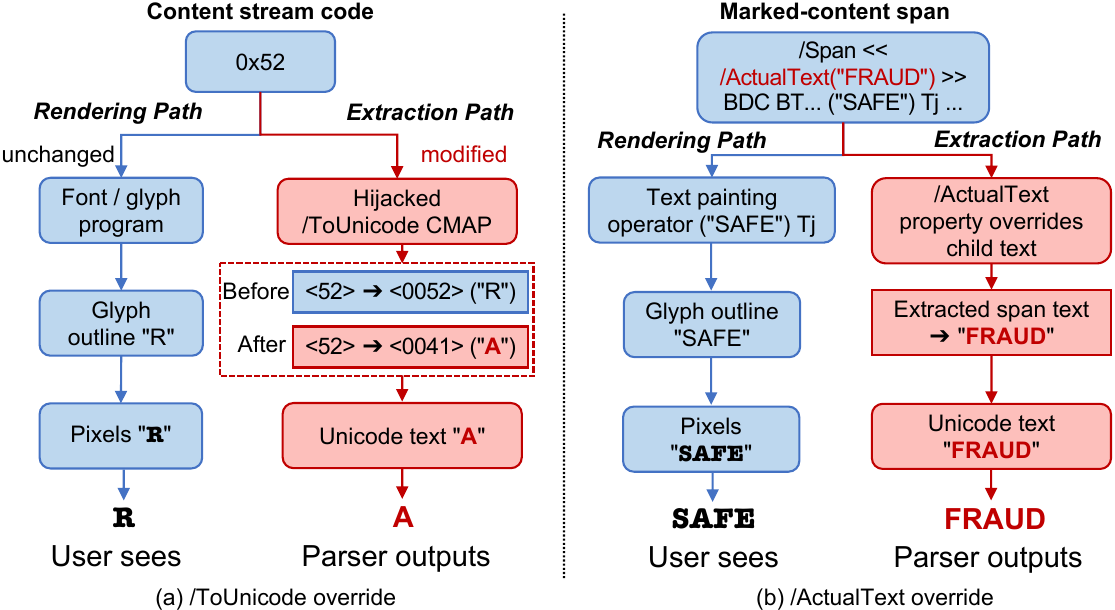}
    \vspace{-5mm}
    \caption{Semantic override mechanisms: rendering follows the font/glyph program while extraction follows a divergent semantic declaration. (a) \texttt{/ToUnicode} CMap remap. (b) \texttt{/ActualText} span substitution.}
    \label{fig:semantic_override}
    \vspace{-4mm}
\end{figure}

\vspace*{2pt}
\noindent \textbf{Semantic Override }
The PDF standard provides two extraction-side mechanisms that can declare
Unicode text independently of the glyphs painted on the page: the
font-level \texttt{/ToUnicode} CMap and the span-level
\texttt{/ActualText} marked-content attribute. A split-view PDF arises
when these declarations describe different text from the visible glyphs.
EG01 targets the font-level channel
(Figure~\ref{fig:semantic_override}(a)). A PDF content stream stores
character codes, not Unicode characters. In Figure~\ref{fig:semantic_override}(a), code
\texttt{0x52} still indexes the font glyph that draws \enquote{R}, but
EG01 changes the corresponding \texttt{/ToUnicode} entry from
\texttt{<0052>} to \texttt{<0041>}. During extraction, the parser follows
this CMap entry, reads \texttt{<0041>}, and emits \enquote{A}; during
rendering, the viewer ignores that Unicode mapping and draws the
unchanged glyph. No extra text object is inserted, and the replacement
applies to every occurrence of the affected code.

EG02 targets the span-level channel
(Figure~\ref{fig:semantic_override}(b)). Marked content can wrap a normal
text operator with \texttt{/ActualText}, originally for accessibility and
copy-paste. 
The renderer paints the child text, such as \texttt{("SAFE")~Tj}, but extractors that honor the wrapper output the declared alternate text, such as \texttt{/ActualText("FRAUD")}, instead.
EG03 applies the same replacement at individual-character granularity. This family requires no
hidden payload layer: the visible page content remains fully visible,
but extraction may be governed
by an explicit semantic channel outside the threat model of existing
hidden-text filters.

\begin{figure}[]
    \centering
    \includegraphics[width=0.9\linewidth]{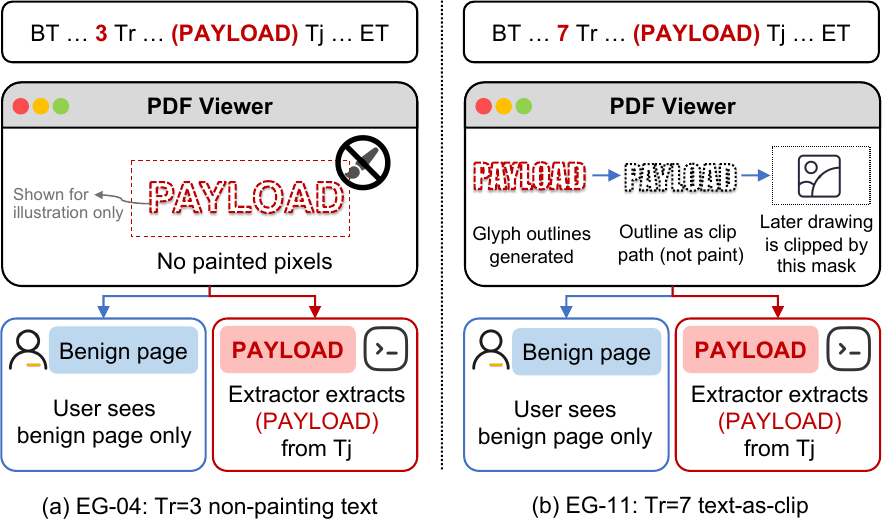}
    \vspace{-2mm}

\caption{Hidden semantic injection via text rendering modes: the payload
remains extractable but is not visible on the rendered page.
(a) EG04 uses \texttt{Tr=3}. (b) EG11 uses \texttt{Tr=7}.}
    \label{fig:semantic_hiddensemantics}
    \vspace{-1mm}
\end{figure}

\vspace*{2pt}
\noindent \textbf{Hidden Semantic Injection }
Hidden semantic injection creates text that is visible to extractors but
invisible in the rendered page. Across this family, hiding can arise from
rendering state, color or transparency, page or clipping geometry, occlusion,
matrix or scale degeneration, or optional-content visibility. The common
pattern is that a human reader sees a benign page, while extractors that read
the PDF content stream may still return the hidden payload.
Appendix~\ref{app:catalog} gives the per-gap details. Below we illustrate two
representative render-mode mechanisms, including EG11, which has no direct
counterpart in prior PDF-to-LLM attack work.

Figure~\ref{fig:semantic_hiddensemantics} illustrates two
text rendering (\texttt{Tr}) modes. EG04 sets \texttt{Tr=3}, the PDF mode that neither
fills nor strokes text: the payload remains in the content stream, but the
renderer paints no glyphs.
EG11 sets \texttt{Tr=7}, which uses the text outline as a clipping path,
i.e., a mask that limits where later drawing can appear, rather than
painting the text itself. In both cases, extractors that decode PDF text
commands without tracking rendering state may return the payload.

\begin{figure}[]
    \centering
    \includegraphics[width=0.9\linewidth]{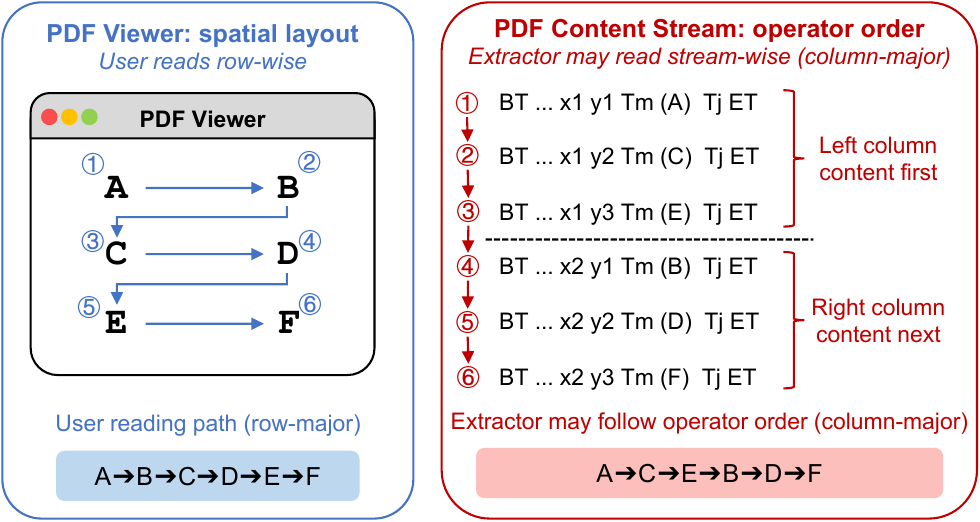}
        \vspace{-2mm}
    \caption{Reading-order split in EG16: two-column visible layout versus column-major content-stream order.}
    \label{fig:family3_readingorder}
    \vspace{-4mm}
\end{figure}


\vspace*{2pt}
\noindent \textbf{Reading-Order Split }
The PDF specification does not require content-stream order to
correspond to spatial reading order. Text fragments are placed on the
page by drawing commands with explicit coordinates, and the order of
those commands in the file often reflects how the PDF was generated,
not how a reader should read the page.
Figure~\ref{fig:family3_readingorder} illustrates EG16, a two-column
layout. The visible page presents six tokens in a row-major grid that a
human reader traverses as
A$\rightarrow$B$\rightarrow$C$\rightarrow$D$\rightarrow$E$\rightarrow$F.
The content stream, however, emits the left column before the right
column. An extractor that follows byte-stream order therefore returns
A$\rightarrow$C$\rightarrow$E$\rightarrow$B$\rightarrow$D$\rightarrow$F,
silently changing cross-column relationships. Extractors that instead
reconstruct reading order from glyph geometry may return the human-visible
sequence, so the same PDF yields different text without hiding any token.
EG17 uses the same construction with three columns, testing the same
stream-order split under a denser layout.


\begin{figure}[]
    \centering
    \includegraphics[width=1\linewidth]{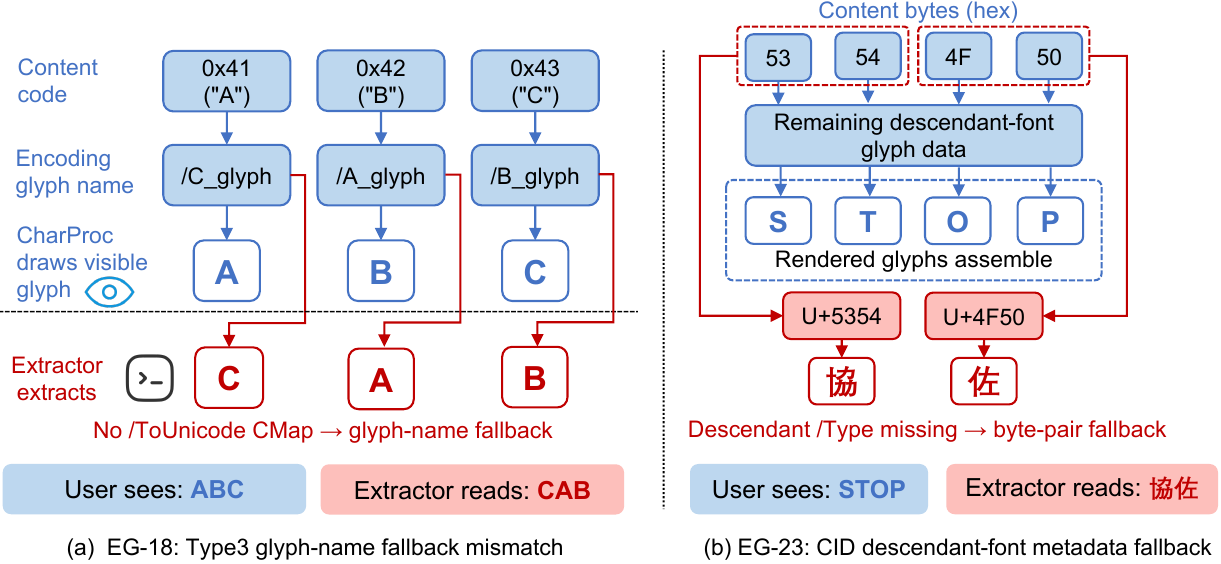}
   \vspace{-4mm}

\caption{Font-decoding split: extractor fallback rules recover text
that differs from the rendered glyphs. (a) EG18 uses Type~3 glyph-name
fallback. (b) EG23 uses CID descendant-font fallback.}
    \label{fig:family4_fontdecoding}
    \vspace{-2mm}
\end{figure}

\vspace*{2pt}
\noindent \textbf{Font-Decoding Split }
For Type~3, composite, and CID-keyed (Character Identifier-keyed) fonts, Unicode recovery can depend
on font metadata and fallback behavior when required fields are absent
or malformed. Figure~\ref{fig:family4_fontdecoding}
illustrates two subfamilies that exploit this hierarchy.
EG18 targets Type~3 fonts
(Figure~\ref{fig:family4_fontdecoding}(a)). A Type~3 font without
\texttt{/ToUnicode} renders each character through a custom
\texttt{CharProcs} glyph program. The encoding table maps content
codes to glyph names in an arbitrary order: code \texttt{0x41} selects
\texttt{/C\_glyph}, code \texttt{0x42} selects \texttt{/A\_glyph}, and
code \texttt{0x43} selects \texttt{/B\_glyph}, so the rendered glyphs
assemble the visible string \texttt{ABC}. An extractor that falls back
to glyph-name lookup in the absence of \texttt{/ToUnicode} recovers
the names in encoding order and returns \texttt{CAB} instead. The
visual and extracted token sequences are permutations of the same
character set, making the divergence invisible to any diff-based
integrity check.
EG23 targets CID composite fonts
(Figure~\ref{fig:family4_fontdecoding}(b)). Removing the descendant
\texttt{/Type} field leaves rendering intact, but some extractors
reinterpret raw content bytes in pairs as two-byte CID or Unicode
values, producing CJK (Chinese, Japanese, and Korean) text codepoints instead of the visible ASCII string.
The broader CID subfamily removes individual metadata fields---\texttt{/Subtype} at the top-level or descendant font,
\texttt{/FontDescriptor}, \texttt{/CIDSystemInfo}, descendant
\texttt{/Type}, or default width \texttt{/DW}---causing extractors
to take different fallback branches. These branch-specific tokens are
stable, making font-decoding splits useful as both exploit carriers and
fingerprinting canaries.

A complete per-mechanism catalog, including precise PDF constructs and mechanisms for each of the
25 confirmed gaps, is provided in Appendix~\ref{app:catalog}
(Table~\ref{tab:app_gap_catalog}).
Table~\ref{tab:prior_work_coverage} separately marks prior-work overlap and
the 14 gaps for which we found no exact or mechanism-level adjacent
construction in the prior work.
We use ``no direct counterpart'' conservatively: a gap qualifies only
when no surveyed prior work instantiates the same PDF path or a
mechanism-level adjacent construction in the same interpretation chain.

\subsection{Specification-Grounded Gap Mining}
\label{sec:gap-mining}

\noindent \textbf{Candidate Mining and Test Generation }
Our mechanism catalog was not assembled through ad hoc PDF tricks. We
mine \emph{candidate render/extract divergence points}: places where
plausible PDF implementation choices can diverge. Because ISO~32000 is prose, we work
over the Arlington PDF Model~\cite{wyatt2021work}, which indexes PDF
objects, keys, values, and constraints in machine-readable tables. An
evidence-constrained LLM-assisted pass proposes candidate divergence
points and bounded mutation spaces from those tables, but these outputs
are treated only as hypotheses; evidence comes from downstream
differential execution and canary validation. Appendix~\ref{app:mining}
provides the prompt and output schema.

We focus on split-view PDF behavior, so we keep only candidates that can
affect visible rendering or extracted text: 
semantic text layers,
visibility and graphics state, page geometry, reading order, and
font-to-Unicode recovery. Candidates outside the render/extract interface
are excluded. We instantiate the retained set with minimal PDF templates
implemented in our generator. Each template places one object class in a
PDF context where it can affect rendering or extraction, such as text
drawn through a font dictionary, painting through a color-space resource,
an invoked XObject, or a mutated page dictionary. For each candidate, the
generator selects the matching template and inserts bounded key/value
variants. The archived generation manifest records 532 executable
divergence test groups, each expanding into multiple bounded PDF variants.
We use cross-engine differential execution as a discovery filter. Each
variant runs over the open-source PDF engines in our parser panel
(Table~\ref{tab:rq1_matrix}), and we compare rendered images and
extracted text strings across engines for the same input. We retain a
group when at least two engines produce distinct outputs: 
visibly distinct rendered images or different normalized text hashes after excluding whitespace-only differences.
This stage retained 154 groups with
observable implementation-level divergence; it does not by itself
establish a split-view security gap. We then focus on the 72 retained
groups with extracted-text divergence, because only these can change the
document text supplied to an LLM pipeline, and reduce them to split-view
candidates in the promotion step below. These groups are concentrated in
CID font mapping, page-geometry and visibility, Type~3 font mapping, and
simple font-dictionary fallback.

\vspace*{2pt}
\noindent\textbf{From Divergences to Confirmed Gaps }
The 72 text-divergent groups are evidence units, not
independent gaps: multiple bounded mutations can induce the
same extractor-output pattern. We collapse them by normalized
extractor output and root PDF mechanism, then instantiate one
representative canary per stable mechanism. A representative is
promoted only if it preserves a stable visible interpretation,
induces non-empty extractor text that diverges from the rendered
view, and is attributable to one dominant PDF mechanism. The
resulting catalog contains 25 confirmed extraction gaps grouped
by security effect in \S\ref{sec:families}. It is not complete
coverage over PDF parser behavior, but a set of robust split-view
primitives validated by deterministic render/extract evidence.
The catalog spans both specification-permitted features
(\texttt{/ToUnicode}, \texttt{/ActualText}, rendering state, and
reading-order freedom) and metadata-incomplete PDFs that robust
renderers tolerate but extractors resolve through fallback paths,
especially in CID font decoding. Both regimes produce the same
security issue: a model-consumed view that differs from the
user-visible page.

\subsection{Two-Tier Benchmark Design}
\label{sec:benchmark}

Our benchmark separates mechanism validation from downstream impact.
Corpus~A tests whether PDF engines traverse a targeted split-view path;
Corpus~B tests whether the resulting extraction divergence propagates
into LLM output. This separation ensures that end-to-end attacks rest on
mechanisms with stable, attributable extractor behavior.

\vspace*{2pt}
\noindent\textbf{Corpus A: Mechanism Canary Corpus }
Corpus~A consists of 25 minimal one-page PDFs, one per confirmed
extraction gap. Each sample isolates one mechanism and embeds a
semantically inert canary token absent from all other corpus documents.
Extractor output is scored by whether it returns the canary, the visible
text, a deterministic artifact, or dropout. Each sample is annotated with
its mechanism family, human-visible content, extractor-side content or
artifact, and expected extractor-level effect.

\vspace*{2pt}
\noindent\textbf{Corpus B: End-to-End LLM Corpus }
Corpus~B provides semantically rich one-page
documents that embed each confirmed gap in a downstream task
carrier rather than a synthetic canary.
Corpus~B uses two task modalities.
\emph{Summary Attack} measures passage-level contamination: semantic
override and hidden semantic injection samples inject a full paragraph
or appendix-style extractor-side claim, and success is scored when the
model repeats it. \emph{QA Attack} measures answer-span corruption: the
mechanism is woven into a short field or sentence, often within the same
line or text object as the visible answer. Success requires the model to
return the extractor-side answer.
Corpus~B covers all 25 extraction gaps. Each gap is assigned one primary
end-to-end modality: 13 are Summary-primary and 12 are QA-primary
(Table~\ref{tab:modality_assignment}). Eleven Summary-primary gaps also
support a clean inline QA carrier (all except EG06 and EG12), adding 11
secondary QA pairs. Thus Corpus~B contains 25 mechanism instances and
36 total gap--modality pairs.

\begin{table}[h]
\centering
\caption{Attack modality assignment. Gaps marked with $\dagger$ also have an inline QA variant.}
\vspace{-2mm}
\label{tab:modality_assignment}
\renewcommand{\arraystretch}{1.25}
\begin{tabular}{ l m{4.5cm} c }
\toprule
Primary Modality & Gaps & Count\\
\midrule
Summary Attack         & EG01$^\dagger$, EG02$^\dagger$,
                         EG04$^\dagger$, EG05$^\dagger$,
                         EG06, EG07$^\dagger$, EG08$^\dagger$,
                         EG09$^\dagger$, EG10$^\dagger$,
                         EG11$^\dagger$, EG12,
                         EG13$^\dagger$, EG14$^\dagger$ & 13\\ \midrule
QA Attack              & EG03, EG15--EG25 & 12\\
\bottomrule
\end{tabular}
\vspace{-1mm}
\end{table}

Each Corpus~B sample records the visible claim or answer span, the
extractor-side content or artifact, and the task prompt. Summary samples
use attacker-chosen claims; QA samples use mutated answers, reordered
spans, or deterministic fallback artifacts. The visible oracle is
established by manual review over the same 10 render-capable PDF tools
used in Table~\ref{tab:rq1_matrix}: a sample is retained only if the
attacker-side claim or answer is not visible in any rendering, and
renderer-dependent benign details are scored from the manually recorded
visible state rather than a single reference renderer.

%% file: pages/evaluation.tex

\section{Evaluation}
\label{sec:evaluation}

In this section, we present our results on various PDF parsers and LLMs,
measuring upstream extraction gaps, downstream model effects, deployment
variation, fingerprinting, and defenses through five research questions:

\begin{enumerate}[leftmargin=3em,topsep=0pt,label=\textbf{RQ\arabic*.}]
    \item How prevalent are extraction gaps across upstream PDF parsers? (\S\ref{sec:eval:upstream})
    \item Do extraction gaps propagate into downstream LLM failures? (\S\ref{sec:eval:downstream})
    \item How do deployment entry points, third-party ingestion layers, and vendor service versions reshape PDF attack risk? (\S\ref{sec:eval:deploymentrisk})
    \item Can canaries we design reliably fingerprint hidden PDF ingestion stacks? (\S\ref{sec:eval:fingerprint})
    \item  How much do existing filters, vision-based ingestion, and our
static scanner reduce split-view risk, and what attack surface remains?
(\S\ref{sec:eval:defense})
\end{enumerate}

\subsection{Settings}

\noindent \textbf{Benchmark Suite }
We evaluate against the two-tier benchmark described in
\S\ref{sec:benchmark}. Corpus~A provides 25 minimal canary
samples for RQ1 and RQ4; Corpus~B provides 36 semantically
rich gap--modality instances for RQ2 and RQ3, with the
Summary/QA assignment shown in Table~\ref{tab:modality_assignment}.

\vspace*{2pt}
\noindent \textbf{PDF Parsers and LLMs }
We evaluate our benchmark against 16 upstream PDF processing stacks: 10 dual-capability PDF parsers that expose both rendering and text extraction, and 6 text-extraction or document-conversion pipelines whose relevant output is the text handed downstream.
Among the 10 dual-capability parsers, three are commercial PDF readers: Adobe Acrobat, Foxit, and PDF Expert. The remaining stacks include open-source libraries and document-processing pipelines commonly deployed in document-to-LLM systems.

For the end-to-end LLM evaluation, we select 7 commercial platforms that expose a native PDF ingestion path through their official API, using the most capable production model with an official PDF-accepting API available in the March~2026 measurement window: Sonnet~4.6, Grok~4.2, GPT-5.4, Gemini~3, Kimi~K2.6, Qwen-Long, and GLM-5.1. Each platform either processes PDF files directly within the model's multimodal context or routes uploads through an officially-documented file extraction chain before inference. All evaluations are conducted via API invocation using each platform's documented PDF submission interface with default generation parameters, ensuring that the preprocessing behavior observed reflects the production ingestion stack rather than any third-party wrapper.
All experiments are conducted on machines running Ubuntu 24.04 with an Intel Core Ultra 9 285K CPU and 64 GB of RAM.

\begin{table*}[]
\centering
\caption{Corpus~A upstream parser outcomes. Left: dual-capability
render/extract consistency; right: text-pipeline exposure.
\redsq\ gap exposed, \greensq\ clean/render-aligned output, \graysq\ dropout/error.}
\label{tab:rq1_matrix}
\label{tab:pure_text_extractors}
\renewcommand{\arraystretch}{1.05}
\vspace{-2mm}
\begin{tabular}{l >{\raggedright\arraybackslash}m{3.0em} @{\hspace{7pt}} *{10}{@{\hspace{1.8pt}}c@{\hspace{1.8pt}}} @{\hspace{5pt}} >{\columncolor{gray!12}}c @{\hspace{10pt}} *{6}{@{\hspace{1.25pt}}c@{\hspace{1.25pt}}} @{\hspace{5pt}} >{\columncolor{gray!12}}c}
\toprule
\multicolumn{2}{c}{} &
\multicolumn{11}{c}{Dual-capability PDF parsers} &
\multicolumn{7}{c}{Text-extraction pipelines} \\
\cmidrule(lr){3-13}\cmidrule(lr){14-20}
\multicolumn{2}{l}{Extraction Gap} &
\rotatebox{90}{MuPDF} &
\rotatebox{90}{Poppler} &
\rotatebox{90}{Xpdf} &
\rotatebox{90}{Pdfium} &
\rotatebox{90}{PDFBox} &
\rotatebox{90}{PDF.js} &
\rotatebox{90}{GhostScript} &
\rotatebox{90}{Adobe Acrobat} &
\rotatebox{90}{PDF Expert} &
\rotatebox{90}{Foxit} &
\cellcolor{white}\rotatebox{90}{Gap exposed} &
\rotatebox{90}{PyPDF} &
\rotatebox{90}{PDFMiner} &
\rotatebox{90}{LlamaParse} &
\rotatebox{90}{OpenDataLoader} &
\rotatebox{90}{Docling} &
\rotatebox{90}{Unstructured} &
\cellcolor{white}\rotatebox{90}{Gap exposed} \\
\midrule

\multirow{3}{*}{\makecell{Semantic\\Override}}
& EG01 & \redsq & \redsq & \redsq & \redsq & \redsq & \redsq & \greensq & \greensq & \redsq & \greensq & 7/10 & \redsq  & \redsq  & \redsq  & \redsq & \redsq & \redsq & 6/6 \\
& EG02 & \redsq & \redsq & \redsq & \redsq & \redsq & \greensq & \greensq & \greensq & \greensq & \greensq & 5/10 & \greensq & \greensq & \greensq & \greensq & \greensq & \greensq & 0/6 \\
& EG03 & \redsq & \redsq & \redsq & \redsq & \redsq & \greensq & \greensq & \greensq & \greensq & \greensq & 5/10 & \greensq & \greensq & \greensq & \greensq & \greensq & \greensq & 0/6 \\
\midrule

\multirow{11}{*}{\makecell{Hidden\\Semantic\\Injection}}
& EG04 & \redsq & \redsq & \redsq & \redsq & \redsq & \redsq & \redsq & \greensq & \redsq & \redsq & 9/10 & \redsq  & \redsq  & \redsq  & \redsq & \redsq & \redsq & 6/6 \\
& EG05 & \redsq & \redsq & \redsq & \redsq & \redsq & \redsq & \redsq & \greensq & \redsq & \redsq & 9/10 & \redsq  & \redsq  & \redsq  & \redsq & \redsq & \redsq & 6/6 \\
& EG06 & \greensq & \greensq & \greensq & \redsq & \redsq & \greensq & \redsq & \greensq & \greensq & \greensq & 3/10 & \redsq  & \redsq  & \greensq & \redsq & \greensq & \redsq & 4/6 \\
& EG07 & \redsq & \redsq & \redsq & \redsq & \redsq & \redsq & \redsq & \redsq & \redsq & \redsq & 10/10 & \redsq  & \redsq  & \greensq & \greensq & \redsq & \redsq & 4/6 \\
& EG08 & \redsq & \redsq & \redsq & \redsq & \redsq & \redsq & \redsq & \redsq & \redsq & \redsq & 10/10 & \redsq  & \redsq  & \redsq  & \redsq & \redsq & \redsq & 6/6 \\
& EG09 & \greensq & \redsq & \redsq & \redsq & \redsq & \redsq & \redsq & \greensq & \redsq & \redsq & 8/10 & \redsq  & \redsq  & \redsq  & \redsq & \redsq & \redsq & 6/6 \\
& EG10 & \redsq & \redsq & \redsq & \redsq & \redsq & \redsq & \redsq & \greensq & \redsq & \redsq & 9/10 & \redsq  & \redsq  & \redsq  & \redsq & \redsq & \redsq & 6/6 \\
& EG11 & \redsq & \redsq & \redsq & \redsq & \redsq & \redsq & \redsq & \greensq & \redsq & \redsq & 9/10 & \redsq  & \redsq  & \redsq  & \redsq & \redsq & \redsq & 6/6 \\
& EG12 & \redsq & \redsq & \redsq & \redsq & \redsq & \redsq & \greensq & \redsq & \redsq & \redsq & 9/10 & \redsq  & \redsq  & \greensq & \greensq & \redsq & \redsq & 4/6 \\
& EG13 & \redsq & \redsq & \redsq & \redsq & \redsq & \redsq & \redsq & \redsq & \redsq & \redsq & 10/10 & \redsq  & \redsq  & \redsq  & \redsq & \redsq & \greensq & 5/6 \\
& EG14 & \greensq & \greensq & \greensq & \redsq & \redsq & \redsq & \greensq & \redsq & \redsq & \redsq & 6/10 & \redsq  & \redsq  & \redsq  & \greensq & \redsq & \redsq & 5/6 \\
\midrule

\multirow{3}{*}{\makecell{Reading-\\Order\\Split}}
& EG15 & \redsq & \greensq & \greensq & \greensq & \redsq & \redsq & \greensq & \greensq & \greensq & \greensq & 3/10 & \redsq  & \greensq & \greensq & \greensq & \greensq & \greensq & 1/6 \\
& EG16 & \redsq & \greensq & \greensq & \redsq & \redsq & \redsq & \greensq & \redsq & \redsq & \greensq & 6/10 & \redsq  & \greensq & \greensq & \redsq & \greensq & \greensq & 2/6 \\
& EG17 & \redsq & \greensq & \greensq & \redsq & \redsq & \redsq & \greensq & \greensq & \redsq & \greensq & 5/10 & \redsq  & \greensq & \greensq & \redsq & \greensq & \redsq & 3/6 \\
\midrule

\multirow{8}{*}{\makecell{Font-\\Decoding\\Split}}
& EG18 & \redsq & \redsq & \graysq & \greensq & \greensq & \greensq & \greensq & \greensq & \greensq & \greensq & 2/10 & \redsq  & \greensq & \greensq & \graysq & \redsq & \greensq & 2/6 \\
& EG19 & \redsq & \greensq & \greensq & \greensq & \greensq & \graysq & \greensq & \redsq & \greensq & \greensq & 2/10 & \greensq & \greensq & \greensq & \graysq & \redsq & \greensq & 1/6 \\
& EG20 & \greensq & \greensq & \greensq & \redsq & \graysq & \graysq & \redsq & \redsq & \redsq & \redsq & 5/10 & \greensq & \greensq & \graysq  & \graysq & \redsq & \greensq & 1/6 \\
& EG21 & \greensq & \redsq & \graysq & \redsq & \graysq & \greensq & \greensq & \graysq & \redsq & \redsq & 4/10 & \redsq  & \redsq  & \graysq  & \graysq & \redsq & \redsq & 4/6 \\
& EG22 & \redsq & \graysq & \greensq & \redsq & \graysq & \greensq & \greensq & \graysq & \redsq & \redsq & 4/10 & \redsq  & \redsq  & \graysq  & \graysq & \graysq & \redsq & 3/6 \\
& EG23 & \redsq & \redsq & \graysq & \redsq & \graysq & \greensq & \greensq & \graysq & \redsq & \redsq & 5/10 & \redsq  & \redsq  & \graysq  & \graysq & \redsq & \redsq & 4/6 \\
& EG24 & \redsq & \redsq & \graysq & \redsq & \redsq & \redsq & \redsq & \redsq & \redsq & \redsq & 9/10 & \redsq  & \redsq  & \redsq  & \graysq & \redsq & \redsq & 5/6 \\
& EG25 & \greensq & \redsq & \graysq & \redsq & \graysq & \greensq & \greensq & \graysq & \redsq & \redsq & 4/10 & \greensq & \redsq  & \greensq & \graysq & \redsq & \redsq & 3/6 \\
\midrule

\multicolumn{2}{c}{Total} & 19/25 & 17/25 & 12/25 & 22/25 & 18/25 & 15/25 & 11/25 & 9/25 & 19/25 & 16/25 & \cellcolor{white}-- & 20/25 & 17/25 & 10/25 & 11/25 & 18/25 & 17/25 & \cellcolor{white}-- \\
\bottomrule
\end{tabular}
\vspace{-1mm}
\end{table*}

\subsection{RQ1: Upstream Extraction Gap Prevalence}
\label{sec:eval:upstream}

Before a model begins reasoning, the text it receives may already diverge
from what the user sees. RQ1 measures this upstream step with Corpus~A
canaries across Table~\ref{tab:rq1_matrix}'s two extractor classes:
dual-capability parsers with rendering and extraction outputs, and
text-extraction pipelines whose observable output is downstream text.
For dual-capability parsers, each cell is checked against that engine's
rendered image: OCR provides a first-pass comparison, and every rendering is
manually inspected. An exposed marker therefore requires benign rendering
plus extraction of the canary or expected artifact. For text-extraction
pipelines, \redsq\ means that this artifact appears in downstream text
without asserting an intra-engine render/extract inconsistency.

\vspace*{2pt}
\noindent\textbf{Intra-Engine Render-Versus-Extract Consistency } 
The left block of Table~\ref{tab:rq1_matrix} reports the intra-engine
consistency matrix for the 10 dual-capability parsers. Across all 25 gaps,
exposed markers dominate, and no parser stays aligned with rendering across the
catalog. 
The key pattern is not only prevalence but structure: tools differ in
whether extraction follows the rendered page, raw text operators, font
metadata, accessibility mappings, or layer visibility.  Since each tool is
compared against its own rendered PNG, a returned marker indicates an internal
render/extract mismatch in that stack, not an artifact of choosing an
overly strict reference viewer.
Conversely, clean markers indicate alignment only
for that mechanism; the same engine often diverges elsewhere.
Even Adobe Acrobat, the most conservative dual parser in our panel,
diverges on 9/25 gaps. At the other end, Pdfium (22/25) exposes nearly
all gaps, while MuPDF and PDF Expert (19/25 each) also expose most of the
catalog.

Semantic overrides show that extractors do not all trust the same
text-to-Unicode source. EG01 poisons \texttt{/ToUnicode}, a mapping used
by many native text extractors, and therefore reaches most stacks. In
contrast, EG02--EG03 use \texttt{/ActualText}, which only affects parsers
that apply marked-content replacement during extraction. Hidden semantic
injection exposes a different split: the payload remains an ordinary
text-showing operand in the content stream, but rendering state prevents
it from appearing on the page. EG07, EG08, and EG13 are therefore exposed
by all dual-capability parsers in our panel. Other cases, such as
no-paint text, clipping, and occlusion, are filtered by only some tools,
showing that visibility checks are selective rather than a general
render/extract consistency guarantee.

Reading-order and font-decoding splits show that a split view does not
require hidden payload text. In reading-order gaps, the visible tokens
remain on the page, but extraction changes their order or binding when it
linearizes the page. In font-decoding gaps, the page still renders the
intended glyphs, but extractors fall back to different text-recovery
rules, producing visible text, anomalous Unicode/CID output, dropout, or
errors. We treat dropout and error outcomes as availability failures, not
as clean extraction-consistency violations; Adobe Acrobat illustrates this
case, as several CID font-decoding canaries fail during text export rather
than returning render-aligned text. The contrast between EG24 (9/10) and
EG18 (2/10) also shows that even the same Type~3 fallback family can
behave differently depending on the concrete canary construction. 
More broadly, the outcome vectors across EG18--EG25 form parser-specific
signatures, motivating their use for parser-stack attribution.

\vspace*{2pt}
\noindent\textbf{Text-Extraction Pipeline Baseline } 
The right block of Table~\ref{tab:rq1_matrix} evaluates six text-extraction or document-conversion pipelines: PyPDF, PDFMiner, LlamaParse, OpenDataLoader, and text-mode Docling and Unstructured. Some may render or OCR internally, but the observable benchmark output is the text that would be handed to a downstream RAG or QA pipeline. An exposed marker therefore means that the pipeline emits the canary or expected artifact into downstream text; it does not assert an intra-engine render/extract inconsistency.

The main result is that extraction pipelines do not turn PDF divergence into
a safer text view. PyPDF and PDFMiner preserve EG01 and every hidden
semantic injection payload, because they recover the PDF text layer rather
than adjudicating human visibility. More sophisticated pipelines are more
selective, but not reliably safer: LlamaParse, OpenDataLoader, Docling, and
Unstructured drop or normalize some difficult font and geometry cases, yet
still expose attack-relevant text across four families. Their dropout/error markers are stable conversion or dropout
signatures, not evidence of semantic consistency.
LlamaParse and OpenDataLoader are the most selective pipelines in the
right block of Table~\ref{tab:rq1_matrix}. LlamaParse exposes the fewest
gaps, but its remaining exposures still span semantic override,
hidden semantic injection, and font-decoding fallback. OpenDataLoader,
which documents rendering-mismatch filters for hidden or visually suppressed
PDF content, suppresses several matching cases in the tested configuration,
but still exposes EG01, several hidden semantic injection gaps, and
reading-order splits EG16--EG17.
Fixed
visibility heuristics can reduce some hidden-text attacks, but they do not
provide semantic consistency against parser differentials.

\begin{tcolorbox}[colback=gray!10, colframe=black, boxrule=0.5pt]
\textbf{Finding I: } \textit{All 25 extraction gaps produce an inconsistent
outcome across the extractor panel. Divergence patterns are stable by
extractor class, consistent with architecture rather than isolated bugs.}
\end{tcolorbox}

\subsection{RQ2: Downstream LLM-Pipeline Exploitability}
\label{sec:eval:downstream}

Upstream extraction gaps become security-relevant only when the
extractor-side document view is consumed by a downstream model. We therefore
evaluate each gap as an end-to-end LLM-pipeline attack using Corpus~B, after
first establishing the same human-visible oracle as \S\ref{sec:benchmark}:
every sample is rendered by the 10 render-capable PDF tools in the left block
of Table~\ref{tab:rq1_matrix} and manually inspected to confirm that the
attacker-side claim, substituted claim, or manufactured answer span is absent
from the rendered pages. 
Corpus~B uses two modalities. \emph{Summary Attack} tests passage-level contamination: the
extractor view contains an attacker-side claim absent from the visible
document, and success requires the response to report that claim rather than
only summarize the visible story. \emph{QA Attack} tests answer-span
corruption: the visible document contains the benign answer, but the extractor
view mutates, appends, reorders, or font-decodes it into an attacker-side span;
success requires returning the extractor-side answer instead of the visible
answer.
We score Summary trials by multi-keyword matching against extractor-side
terms absent from the visible content; QA trials use exact answer matching,
and each trial is a separate API call.
As a matched control, we remove the gap construct from every Corpus~B sample while keeping the visible story and prompt unchanged; across 1,260 matched-control runs (36 gap–modality pairs × 7 platforms × 5 trials), none produces the attacker-side claim or answer.
Table~\ref{tab:modality_assignment} lists the full assignment;
Appendix~\ref{app:rq2-prompts} gives the benign prompt templates and
representative visible/extractor-side carriers. (Figure~\ref{fig:app_eg02_example}, Figure~\ref{fig:app_eg04_example}, and Figure~\ref{fig:app_eg11_qa_example}).

\begin{figure}[]
    \centering
    \includegraphics[width=1\linewidth]{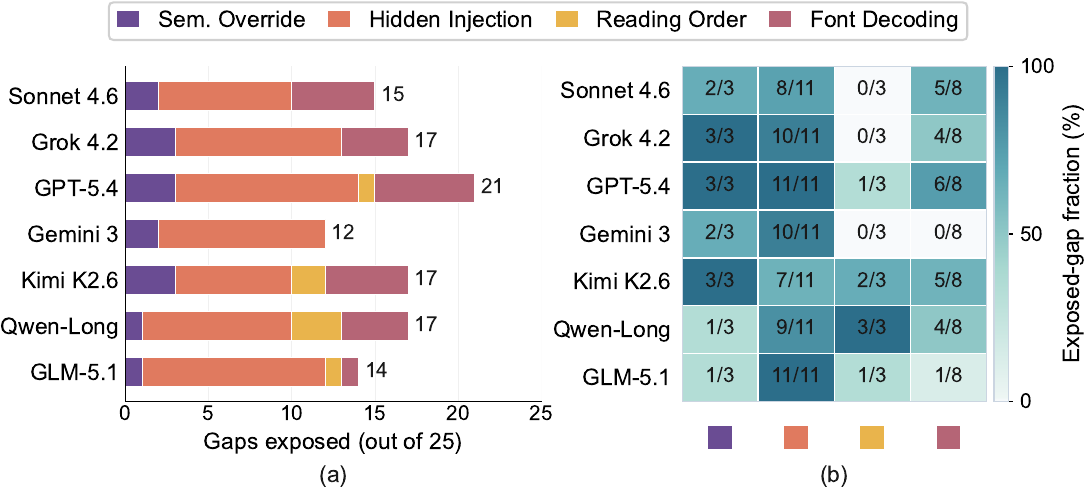}
    \vspace{-6mm}
    \caption{Per-platform extraction-gap coverage by 2amily. (a) Exposed-gap counts, stacked by family. (b) Family-level exposed-gap fractions.}
    \vspace{-5mm}
    \label{fig:rq2_bar}
\end{figure}

\begin{figure}[]
    \centering
    \includegraphics[width=1\linewidth]{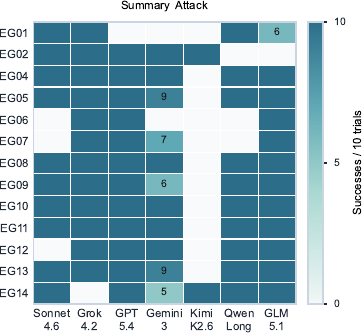}
    \vspace{-5mm}
    \caption{Summary attack heatmap over the 13 Summary-primary gaps. Each cell reports successes out of 10 trials for a given gap–platform cell. Annotated values mark partial scores (1--9); blank cells indicate 0/10.}
    \label{fig:rq2_summary_heatmap}
    \vspace{-2mm}
\end{figure}

\begin{figure}[]
    \centering
    \includegraphics[width=1\linewidth]{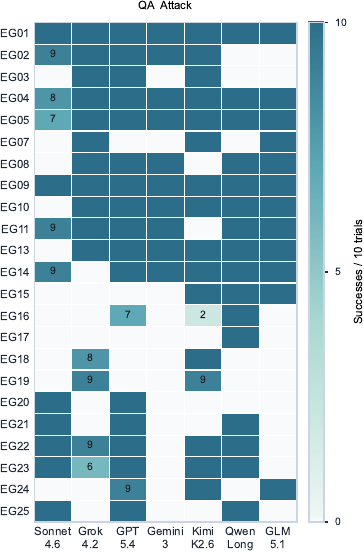}
    \vspace{-5mm}
    \caption{QA attack heatmap over the 12 QA-primary gaps and the 11 cross-modality QA variants of Summary-primary gaps.}
    \label{fig:rq2_qa_heatmap}
    \vspace{-4mm}
\end{figure}

\vspace*{2pt}
\noindent\textbf{Overall Gap Coverage }
Figure~\ref{fig:rq2_bar} summarizes EG-level coverage: an
EG--platform pair is counted once if any evaluated attack form succeeds
in at least one of ten independent API calls. Every platform is affected
under this criterion, with coverage ranging from 12/25 to 21/25 gaps.
The repeat matrices (Figures~\ref{fig:rq2_summary_heatmap}
and~\ref{fig:rq2_qa_heatmap}) show that this coverage is usually stable
rather than one-off: under a stricter high-repeat cutoff, 65/69 exposed
Summary EG--platform pairs and 96/100 exposed QA EG--platform pairs
succeed in at least 8/10 trials.
Treating the calls as Bernoulli trials, two-sided Wilson 95\% confidence intervals 
remain broad at \(n=10\) (10/10: 72–100\%; 8/10: 49–94\%), so \(k/10\) measures repeatability 
rather than a precise per-call success probability.
We manually audited every successful Summary output. We count success only
when the model treats the extractor-side claim as document content, not
when it merely warns about hidden or non-visible text.

\begin{tcolorbox}[colback=gray!10, colframe=black, boxrule=0.5pt]
\textbf{Finding II: } \textit{In our benchmark, downstream failures are primarily induced upstream: when the ingestion stack admits the extractor-side view, LLMs usually treat it as ordinary document evidence.}
\end{tcolorbox}

\vspace*{2pt}
\noindent\textbf{Attack Modality Matters }
Figures~\ref{fig:rq2_summary_heatmap} and~\ref{fig:rq2_qa_heatmap} separate two risks hidden by a single coverage number. Summary attacks test whether hidden or overridden passage-level content contaminates free-form generation, making them strongest for semantic override and hidden semantic injection. Many summary cells are stable `10/10', showing that once injected passages enter document text, models often summarize them as ordinary content.
QA attacks expose a different failure mode: models become faithful readers of the wrong extracted span. This is why QA is essential for reading-order split and font-decoding split. These families do not necessarily create a fluent hidden paragraph, but they can corrupt the binding between a question and its answer, or mutate a visible token through font fallback. The QA heatmap therefore captures exact-answer failures that summary evaluation would understate. In particular, reading-order gaps are sparse but highly diagnostic, while font-decoding gaps produce platform-specific corrupted spans that reveal the model is consuming a non-visual decoding path.

\vspace*{2pt}
\noindent\textbf{Pipeline Architecture Shapes the Profile }
The most informative axis is not the model name alone but the upstream 
PDF ingestion architecture. Native PDF platforms and adapter-mediated 
platforms expose different subsets of the same gap catalog. 
GPT-5.4 has the broadest coverage (`21/25'), spanning all semantic 
override and hidden semantic injection gaps, one reading-order gap, 
and most font-decoding gaps. Gemini~3 has the lowest total (`12/25'), 
but its failures are not random: it is highly vulnerable to hidden 
semantic injection (`10/11') while showing no successful font-decoding attacks. This profile is consistent with an ingestion path that admits many hidden text objects but does not expose the same composite-font fallback behavior.

The adapter-mediated systems show even clearer parser fingerprints. Qwen-long is exposed to all three reading-order gaps, indicating a content-order or file-extraction path that preserves operator order where a visual reader would use spatial ordering. GLM-5.1 is exposed to all hidden semantic injection gaps but just one of the font-decoding family, suggesting a text extractor that retains hidden text but normalizes or drops harder font fallback cases. Kimi~K2.6 is more selective: it is exposed to all semantic override gaps and many font-decoding and reading-order gaps, but only `7/11' hidden semantic injection gaps. This mixture indicates that platform-side filtering can suppress some hidden-content carriers without providing semantic consistency against override, ordering, or font-decoding differentials.

\begin{tcolorbox}[colback=gray!10, colframe=black, boxrule=0.5pt]
\textbf{Finding III: } \textit{Exposure is stable and platform-specific: most heatmap cells are either 0/10 or high-repeat successes, with only a small number of partial outcomes, so each platform largely admits or drops a given extraction gap consistently across trials.}
\end{tcolorbox}

\vspace*{2pt}
\noindent\textbf{Beyond Hidden Text }
Figure~\ref{fig:rq2_bar}(b) separates exposed-gap coverage by mechanism family.
Hidden semantic injection is the broadest carrier because a pipeline need
only retain renderer-suppressed text. The more important result is that
exposure persists outside hidden text. Semantic override changes the
Unicode authority used by extraction; reading-order split changes the
linearization of visible tokens; and font-decoding split exposes
extractor-specific fallback artifacts. These families do not require an
invisible paragraph or hidden instruction, yet they still cause models to
summarize or answer from a document view the user did not see.
\subsection{RQ3: Deployment-Layer PDF Risk}
\label{sec:eval:deploymentrisk}

RQ2 treats each platform as a single PDF-to-LLM pipeline. RQ3 varies the
production path that constructs the model's document context: hosted cloud
routes, Web chat products, vendor service versions, and local client
applications. These paths are not thin wrappers around the same model. Each
can introduce its own PDF renderer, parser, OCR path, file adapter, chunker,
or safety filter before the model sees the document. All deployment paths
follow the RQ2 unique-EG rule: an EG is
counted as exposed if any evaluated attack form succeeds at least once
under that path, with Web-chat and local-runtime paths evaluated manually.

\vspace*{2pt}
\noindent\textbf{Alternative Cloud Deployments }
Table~\ref{tab:cloud_backend} compares official API access with
provider-restricted cloud routes for the same nominal model. The two
routes show different effects. GPT-5.4 through Azure drops from 21/25
to 17/25. The missing gaps are concentrated in geometry-sensitive cases:
off-page positioning (EG06), tiny text (EG07), zero-height text matrices
(EG12), and a reading-order split (EG16). In contrast, semantic override
and font-decoding gaps largely persist. This pattern suggests that Azure's hosted route
changes the PDF preprocessing boundary: it appears to normalize or filter
some spatially degenerate text, but does not remove semantic remapping or
font-metadata divergences.

\begin{table}[h]
\vspace{-3mm}
\centering
\caption{Official-API vs. cloud-backend PDF exposure for models tested on both routes.}
\label{tab:cloud_backend}
\renewcommand{\arraystretch}{1.2}
\begin{tabular}{ l c c c }
\toprule
LLM & Official API & Cloud Backend & $\Delta$ \\
\midrule
GPT-5.4 (Azure)  & 21 & 17 & $\downarrow$4 \\
Sonnet 4.6 (Bedrock) & 15 & 15 & 0\\
\bottomrule
\end{tabular}
\end{table}

Sonnet 4.6 through Amazon Bedrock remains at 15/25, matching the official
API count. This does not mean that Bedrock is a neutral transport layer in
general; it means that, for this model and corpus, the routed deployment
does not measurably reduce aggregate exposure. The contrast with Azure is
the important point: changing the hosting path can change which PDF
ambiguities reach the model, even when the nominal model is unchanged.

\begin{tcolorbox}[colback=gray!10, colframe=black, boxrule=0.5pt]
\textbf{Finding IV: } \textit{Cloud deployment is a PDF security boundary. The same nominal model can expose a different gap set when routed through a hosted backend, because the backend may change the parser, converter, or normalization layer before inference.}
\end{tcolorbox}

\noindent\textbf{Product Entry Points: API versus Web Chat }
Table~\ref{tab:webchat} compares official API access with each vendor's
Web chat product. The direction is not uniform: Web chat is sometimes
less exposed, sometimes more exposed, and sometimes unchanged. This
already rules out treating a vendor's API and Web product as equivalent
security surfaces.
Gemini is the clearest example. Its API path exposes 12/25
gaps, while its Web chat path is 0/25 on the one-page corpus. The affected
API gaps span semantic override, hidden semantic injection, and
font-decoding split, all of which require a text extraction view to reach
the model. The Web result is consistent with a rendering- or vision-first
frontend that does not expose the extractor-side text layer on these
short documents. This is not a semantic guarantee, however: in a page-count
probe, 40- and 80-page variants remain at 0/25, but a 120-page version with
the attack page at the end falls back to a text-oriented route and exposes
21/25 gaps.
Thus, the apparent protection is a routing heuristic, not a robust PDF
consistency defense.

\begin{table}[h]
\centering
\caption{Exposed gap counts via official API and Web chat.}
\label{tab:webchat}
\renewcommand{\arraystretch}{1.2}
\begin{tabular}{ l c c c }
\toprule
LLM & Official API & Web Chat & $\Delta$ \\
\midrule
Sonnet 4.6  & 15 & 19 & $\uparrow$4  \\
Grok 4.2    & 17 & 18 & $\uparrow$1  \\
GPT-5.4     & 21 & 19 & $\downarrow$2 \\
Gemini 3    & 12 & 0  & $\downarrow$12 \\
Kimi K2.6   & 17 & 14 & $\downarrow$3 \\
Qwen 3.5$^\dagger$ & 17 & 13 & $\downarrow$4 \\
GLM-5.1     & 14 & 14 & 0  \\
\bottomrule
\multicolumn{4}{l}{\scriptsize{$\dagger$ Qwen-Long is API-only; Web chat test uses Qwen~3.5.}}\\
\end{tabular}
\end{table}

Other products move in the opposite direction. Sonnet 4.6 and Grok 4.2
are more exposed through Web chat than through API access, indicating
that their Web frontends admit additional PDF-derived text views not
present in the API path. GPT-5.4, Kimi K2.6, and Qwen show reduced Web
coverage, while GLM-5.1 is unchanged. Qwen is a special case because the
API result uses Qwen-Long's dedicated file-processing path, whereas the
Web experiment uses Qwen~3.5; we therefore interpret it as a vendor
entry-point comparison rather than a strict same-model comparison.

\begin{tcolorbox}[colback=gray!10, colframe=black, boxrule=0.5pt]
\textbf{Finding V: } \textit{API and Web chat are distinct PDF pipelines, not interchangeable access modes. A Web frontend can eliminate, preserve, or introduce exposed gaps depending on its rendering, OCR, file-conversion, and routing heuristics.}
\end{tcolorbox}

\vspace*{2pt}
\noindent\textbf{Within-Vendor Version Updates }
Model-version labels can hide ingestion-service changes. 
We therefore extend Corpus~B to additional OpenAI and Anthropic
versions while keeping the main RQ2 table fixed. The extension uses the
same prompts, submission path, and EG-level success criterion as RQ2, and
Table~\ref{tab:within_vendor} reports only EG-level coverage.

\begin{table}[!h]
\centering
\caption{Within-vendor model version comparison on Corpus~B. Main RQ2
rows are included as references.}
\vspace{-1.5mm}
\label{tab:within_vendor}
\renewcommand{\arraystretch}{1.15}
\begin{tabular}{ l l c c }
\toprule
Vendor & Model & Exposed EGs & $\Delta$ \\
\midrule
\multirow{3}{*}{OpenAI}
& GPT-5.3 & 21/25 & 0 \\
& GPT-5.4 (main) & 21/25 & ref. \\
& GPT-5.5 & 18/25 & $\downarrow$3 \\
\midrule
\multirow{3}{*}{Anthropic}
& Haiku~4.5  & 15/25 & 0 \\
& Sonnet~4.6 (main) & 15/25 & ref. \\
& Opus~4.7   & 15/25 & 0 \\
\bottomrule
\end{tabular}
\end{table}

The two vendors show different update patterns. Anthropic is internally
stable: Haiku~4.5, Sonnet~4.6, and Opus~4.7 expose the same 15/25 gaps with
the same per-EG pattern, despite spanning different capability tiers. This
indicates that the shared ingestion service, rather than model capability,
dominates exposure. OpenAI shows a narrow service update. GPT-5.3 and
GPT-5.4 expose the same 21/25 gaps, while GPT-5.5 drops to 18/25 by no
longer exposing EG06 (off-page text), EG07 (near-zero font size), and
EG12 (zero-height text matrix). These three are degenerate geometric
hiding constructions. Semantic override, most hidden-injection mechanisms,
and font-decoding gaps persist across all three GPT versions.
This pattern echoes the Azure-deployed GPT-5.4 result above
(Table~\ref{tab:cloud_backend}): two independent ingestion-side changes,
a third-party cloud route and a vendor version update, converge on narrow
geometric filtering while leaving the structural split-view class intact.

\vspace*{2pt}
\noindent\textbf{Local Deployment }
Ollama and LM Studio, two widely used local model applications with built-in
PDF interaction, expose a third deployment pattern. PDF parsing happens
inside the client application before any tokens reach the local model. We
therefore evaluate this setting at the
ingestion boundary rather than as a benchmark of local model capability:
for each sample, we inspect the model input context and count exposure
when the extractor-side content is present in that context. This design is
intentional. The local model is constrained by commodity hardware and is
not the object of attribution; once a hidden or corrupted PDF view has
entered the prompt, whether a model follows, refuses, or recognizes it as
suspicious is a downstream prompt-injection robustness question outside
the scope of this RQ.

Table~\ref{tab:local} shows that local deployment does not remove the
attack surface. Ollama exposes extractor-side content for 16/25 gaps
and LM Studio for 17/25, comparable to the exposure rates of commercial
API paths in RQ2. The reason is structural: the local model receives
whatever text the desktop application's PDF extractor emits. Hidden text
operands, off-page text, and reordered spans can therefore enter the
local context without any cloud-side ingestion filter.

\begin{table}[h]
\vspace{-3mm}
\centering
\caption{Local deployment ingestion outcomes. Timeout marks parser or
runtime failure before usable context.}
\vspace{-1.5mm}
\label{tab:local}
\renewcommand{\arraystretch}{1.2}
\begin{tabular}{ l c c }
\toprule
 & Ollama & LM Studio \\
\midrule
Gap exposed     & 16 & 17 \\
Not exposed     & 9  & 5  \\
Timeout         & 0  & 3  \\
\bottomrule
\end{tabular}
\end{table}

The non-exposed cases are also informative. EG02 and EG03 do not
propagate on either local runtime because the PDF extractors bundled with
Ollama and LM Studio ignore \texttt{/ActualText}; they return the visible
stream text rather than the override. Thus, local resistance is not a
model property, but a parser capability mismatch. LM Studio additionally
times out on EG21, EG22,
and EG25. These cases produce anomalous font-decoding artifacts in the
input context; the result is not a successful payload injection, but an
availability failure caused by parser output that the local runtime does
not handle gracefully.

\begin{tcolorbox}[colback=gray!10, colframe=black, boxrule=0.5pt]
\textbf{Finding VI: } \textit{Local deployment shifts PDF risk to the client-side extractor. Without a managed ingestion service, parser artifacts can directly become model context, and in some font-decoding cases even trigger availability failures.}
\end{tcolorbox}

\subsection{RQ4: Ingestion-Stack Fingerprinting}
\label{sec:eval:fingerprint}

RQ1 shows that the same PDF can yield different extractor-side strings
across parsers. RQ4 asks whether these differences are stable enough to
fingerprint the hidden preprocessing path behind a PDF-to-LLM pipeline.
The key observation is that extraction gaps are not only attack vectors:
they are also diagnostic probes. A gap that causes one parser to emit a
glyph name, another to emit a Unicode fallback, and another to drop the
span entirely creates a compact signature of the ingestion stack.

\vspace*{2pt}
\noindent\textbf{Controlled Parser-Panel Fingerprints }
We first validate the idea on the controlled extractor panel from RQ1.
We construct a packed four-page canary from four main gaps: EG01
(\texttt{/ToUnicode} semantic override), EG17 (reading-order split),
EG19 (font-decoding fallback), and EG25 (CID fallback). This canary spans
three mechanism families, using two font-decoding probes to exercise
different fallback paths, and produces a tuple of short extracted strings.
After the same canonicalization used by the fingerprint oracle, the tuple
uniquely separates all 16 parser and ingestion paths in the RQ1 panel,
including dual-capability parsers and text-extraction pipelines.
This result does not rely on a single fragile parser quirk. The four
pages exercise independent axes of PDF processing: semantic mapping,
layout linearization, font fallback, and CID recovery. A pipeline may
agree with another pipeline on one axis, but disagreement on any other
axis separates their signatures. This is why a small canary set can
distinguish a heterogeneous parser panel.

\begin{tcolorbox}[colback=gray!10, colframe=black, boxrule=0.5pt]
\textbf{Finding VII: } \textit{Extraction gaps are diagnostic as well as attack-relevant. A four-page canary spanning semantic mapping, layout order, and font fallback uniquely separates all 16 parser paths in our controlled panel.}
\end{tcolorbox}

\vspace*{2pt}
\noindent\textbf{Framework-Level Fingerprints }
We next test whether the same principle applies to realistic LLM framework configurations. 
We use LangChain~\cite{langchain_document_loaders}, a widely used LLM
application framework with multiple PDF loaders that wrap different
underlying extraction stacks.
LangChain is parser-agnostic: the security
behavior of a LangChain pipeline depends on which PDF loader is selected,
not on LangChain as a framework. We therefore configure seven concrete
LangChain PDF loaders: PyPDFLoader, PyMuPDFLoader, PyPDFium2Loader,
PDFMinerLoader, DoclingLoader in text mode, OpenDataLoaderPDFLoader,
and Unstructured in fast mode. We exclude PDFPlumberLoader from
the independent target set because it is a PDFMiner-family wrapper.

Rather than using the packed four-page canary, we search for a minimal
set of individual PDF probes that separates these loader backends. The
selected set contains two documents: an EG12 zero-height-matrix suffix
carrier, which separates loaders by whether the hidden suffix is joined,
displaced, partially recovered, or removed, and EG18, which separates
Type~3 glyph-name fallback behavior. Together, these two documents
cover all 21 pairwise distinctions among the seven loaders with no
unresolved collisions.

\begin{table}[]
\centering
\caption{LangChain loader signatures for the two-document fingerprint set. Entries show the salient normalized output used for attribution.}
\vspace{-1.5mm}
\label{tab:langchain_signatures}
\renewcommand{\arraystretch}{1.2}
\resizebox{0.477\textwidth}{!}{
\begin{tabular}{ l l l }
\toprule
Loader & EG12 & EG18 \\
\midrule
PyPDFLoader        & \texttt{LARK-27XQ9}     & \texttt{/C\_GLYPH/A\_GLYPH/B\_GLYPH} \\
PyMuPDFLoader      & \texttt{LARK-27XQ9}     & \texttt{CAB} \\
PyPDFium2Loader    & \texttt{LARK-27XQ9}     & \texttt{ABC} \\
PDFMinerLoader     & \texttt{LARK-27X...Q9}  & \texttt{ABC} \\
DoclingLoader      & \texttt{XQ9...LARK-27}  & \texttt{/C\_GLYPH/A\_GLYPH/B\_GLYPH} \\
OpenDataLoader     & \texttt{LARK-27}        & \texttt{EMPTY} \\
Unstructured       & \texttt{LARK-27X}       & \texttt{ABC} \\
\bottomrule
\end{tabular}
}
\vspace{-1mm}
\end{table}

The table illustrates why multi-probe fingerprints are necessary.
EG12 alone cannot distinguish PyPDF, PyMuPDF, and PyPDFium2, which all
recover the full suffix \texttt{LARK-27XQ9}. EG18 alone cannot
distinguish PyPDFium2, PDFMiner, and Unstructured, which all recover the
visible token \texttt{ABC}. The pair is discriminative because the two
mechanisms fail along different parser boundaries. Thus, fingerprinting
does not require a large probe corpus; it requires probes whose
ambiguities are orthogonal.

\vspace*{2pt}
\noindent\textbf{Scope of Attribution }
In the controlled settings above, where parser or loader identities are
known, the canaries separate all 16 parser paths and all seven LangChain
loader backends. In black-box production deployments, however, an output
tuple should be interpreted as family-level or behavior-level evidence
rather than exact parser-version attribution. Wrappers, version drift,
OCR fallbacks, and platform-side cleanup or chunking can collapse or
modify signatures. The security value is therefore not exact library
identification, but making hidden preprocessing behavior observable: it
tells an auditor which PDF view is likely to reach the model.

\subsection{RQ5: Defense Evaluation}
\label{sec:eval:defense}

We evaluate three classes of defense against extraction-gap attacks:
existing extraction-layer safety filters, vision-based processing, and a
proposed static scanner. For each defense, we focus on what class of PDF
divergence it actually controls, and where the protection fails.

\vspace*{2pt}
\noindent\textbf{Existing Safety Filters }
OpenDataLoader is the only publicly available PDF extractor in our panel
that advertises built-in PDF prompt-injection safety filters. These
filters should be understood as fixed rendering-mismatch heuristics, not
as a semantic consistency checker. In the tested configuration,
OpenDataLoader suppresses mechanisms that match its rule set, such as
near-zero text size (EG07), zero-height text (EG12), and hidden
optional-content layers (EG14). It still exposes many other hidden-state
variants (EG04--EG06, EG08--EG11, EG13), as well as semantic override,
reading-order, and font-decoding gaps. The failure mode is therefore a
scope mismatch: visibility heuristics can remove some hidden chunks, but
they do not enforce semantic consistency between the rendered page and
the extracted text.

Kimi K2.6 shows a related but more deployment-specific filtering pattern.
Its file-extract path largely suppresses standalone hidden paragraphs in
the one-page summary carriers, but the same mechanisms remain effective
when encoded at a smaller granularity. In the QA carriers, the visible
field contains \texttt{Review code: LARK-27}, while the extractor-side
answer is formed by an inline hidden suffix such as \texttt{XQ9} placed in
the same line or text object. Kimi remains exposed on most corresponding
QA gaps, showing that chunk-level filtering does not provide
mechanism-complete protection.

\begin{tcolorbox}[colback=gray!10, colframe=black, boxrule=0.5pt]
\textbf{Finding VIII: } \textit{Existing PDF safety filters cover only a limited subset of hidden-text constructions. They can suppress some standalone hidden chunks, but do not address semantic overriding.}
\end{tcolorbox}

\vspace*{2pt}
\noindent\textbf{Render-Based Defense and Its Deployment Limits }
Render-based processing is the strongest defense against text-layer
split-view attacks when it is applied consistently. If a platform renders
each PDF page to an image and extracts content from the visual
representation, extractor-only text that is not visible to the user
cannot directly enter the model context. RQ3 observes behavior consistent
with this defense in the Gemini~3 Web Chat setting: one-page submissions
achieve 0/25 attack successes, versus 12/25 through Gemini's API
extraction path.

\begin{table}[h]
\vspace{-2mm}
\centering
\caption{Gemini 3 Web Chat attack success under page-count scaling, broken down by family.}
  \vspace{-2mm}
\label{tab:gemini_ocr_bypass}
\renewcommand{\arraystretch}{1.2}
\begin{tabular}{ l l l l l l }
\toprule
Pages & Sem. & Hidden & Reading & Font & Total \\
\midrule
1   & 0/3 & 0/11 & 0/3 & 0/8 & 0/25 \\
40  & 0/3 & 0/11 & 0/3 & 0/8 & 0/25 \\
80  & 0/3 & 0/11 & 0/3 & 0/8 & 0/25 \\
120 & 3/3\,\textcolor{red!75!black}{$\uparrow$} &
11/11\,\textcolor{red!75!black}{$\uparrow$} &
1/3\,\textcolor{red!75!black}{$\uparrow$} &
6/8\,\textcolor{red!75!black}{$\uparrow$} &
21/25\,\textcolor{red!75!black}{$\uparrow$} \\
\bottomrule
\end{tabular}
\end{table}

A likely deployment pressure is cost.
Rendering and vision processing scale with page count, so a cost-sensitive
routing change could explain different behavior on long documents.
We probe this behavior by padding each
submission with benign filler pages while keeping the attack payload on
the final page. Gemini 3 Web Chat remains at 0/25 through 1-, 40-, and
80-page submissions, but changes to 21/25 at 120 pages
(Table~\ref{tab:gemini_ocr_bypass}). Semantic-override and hidden-injection
attacks then succeed completely, while reading-order and font-decoding show
partial resistance. Thus, the results establish that the protection observed for
shorter documents does not persist at 120 pages. The user receives no indication 
that the ingestion mode has changed. 
This makes render-based processing an effective defense only when the
platform commits to visual processing across document sizes, rather than
using it as a short-document optimization.

\begin{tcolorbox}[colback=gray!10, colframe=black, boxrule=0.5pt]
\textbf{Finding IX: } \textit{In Gemini Web Chat, document length changes
the observed ingestion behavior: shorter submissions are consistent with
render-based ingestion, whereas at 120 pages its behavior becomes
consistent with text extraction, re-exposing the model to extractor-side
content.}
\end{tcolorbox}

\vspace*{2pt}
\noindent\textbf{Proposed Static Scanner }
We implement a lightweight scanner that flags structural conditions associated with our extraction gaps before a document is sent to an LLM. The scanner parses PDF objects, content streams, text-rendering state, marked-content attributes, optional-content groups, and font metadata. It
does not search for a particular payload string; instead, it checks for
mechanisms that can separate the rendered view from the extracted view,
such as extractor-side replacement text, non-painted text operators,
disabled layers, reading-order ambiguity, and incomplete font-to-Unicode
metadata. Each rule triggers on the Corpus~A canary for the gap it was
designed to detect. 
We evaluate runtime and benign finding rates on two corpora:: 5{,}000 Web PDFs from
the PDF Association corpus~\cite{pdf_corpora}, and 4{,}722 public
academic papers from recent security and AI venues, with venue
distribution in Appendix~\ref{app:scanner-sampling}. 
On the full academic corpus, the scanner averages under 0.3\,s/PDF, whereas a
Poppler-render-plus-Tesseract-OCR baseline averages over 20\,s/PDF on the
same hardware. 
We report
\emph{any finding}, where a document triggers at least one family rule,
and \emph{high-confidence finding}, where the matched structure directly
changes extractor-visible text or closely matches a confirmed mechanism.
High confidence is a structural signal, not a claim that the document is
malicious.

\begin{table}[h]
\vspace{-2mm}
\centering
\caption{Scanner benign finding rates by family. Rates are document-level;
families are non-exclusive.}
\label{tab:scanner_benign_family}
\vspace{-2mm}
\scriptsize
\setlength{\tabcolsep}{1.8pt}
\renewcommand{\arraystretch}{1.12}
\begin{tabular}{lcccc}
\toprule
 & \multicolumn{2}{c}{Web} &
   \multicolumn{2}{c}{Academic} \\
\cmidrule(lr){2-3}\cmidrule(lr){4-5}
Family & Any & High-conf. & Any & High-conf. \\
\midrule
Semantic override & 0.46\% & 0.46\% & 0.02\% & 0.02\% \\
Hidden injection  & 0.48\% & 0.28\% & 0.11\% & 0.04\% \\
Reading order     & 1.53\% & 0.00\% & 3.08\% & 0.00\% \\
Font decoding     & 0.24\% & 0.24\% & 6.81\% & 6.81\% \\
\bottomrule
\end{tabular}
\vspace{-2mm}
\end{table}

Table~\ref{tab:scanner_benign_family} shows the intended calibration.
Semantic-override and hidden-injection findings are rare in both benign
corpora, so the strongest claim-rewriting and hidden-text mechanisms impose
low benign burden. Reading-order ambiguity is more common, but never high
confidence, because multi-column layouts and tables are ordinary document
structures. Font-decoding findings are elevated in academic PDFs because
LaTeX-derived toolchains often embed Type~3 or subset fonts without
authoritative \texttt{/ToUnicode} maps: these files render correctly, but
extractors must recover Unicode through fallback heuristics. To validate
this interpretation, we manually inspected 30 randomly sampled
high-confidence font-decoding findings; all were benign
\texttt{FONT\_TYPE3\_NO\_TOUNICODE} hits, mostly from TeX-derived academic
PDFs, and none contained attack-like payloads. We therefore treat
font-decoding findings as extraction-integrity audit signals requiring
render/extract confirmation, not as standalone blocking evidence.

We also cross-check PDF-Prompt-Injection-Toolkit
\cite{pdfpromptinjectiontoolkit}, an open-source red/blue-team toolkit for
PDF-to-LLM prompt injection. Its generator covers common hidden prompt
carriers such as white text, tiny text, off-page text, OCG layers,
metadata, and zero-width Unicode; at the attack-generation level, this
overlaps four gaps in our catalog (EG05, EG06, EG07, and EG14). Our static
scanner flags 5 of 7 toolkit samples; the two misses, metadata and
zero-width Unicode, are outside our 25-gap render/extract scope. 
In the reverse direction, the toolkit scanner flags EG05, EG06, EG07,
EG09, EG12, and EG14 through hidden-text heuristics; this shows existing
detection coverage, not prior attack mechanisms.
The toolkit
does not target semantic override, reading order, or font-decoding gaps.
Thus, the scanner is best used for intake triage: it flags PDFs that
warrant render/extract confirmation, not standalone blocking.

%% file: pages/discussion.tex
\section{Discussion}

\noindent\textbf{Semantic Integrity Requires a Boundary Contract }
For LLM ingestion, the platform must enforce a contract between the
user-visible meaning and the model-consumed text. Patching individual
extractors can narrow specific mechanisms, but new compliant
implementations may re-expose the same ambiguity. RQ3 shows this pattern
in deployed services: GPT-5.5 removes three geometric-hiding gaps while
preserving semantic-override and font-decoding exposure, whereas
Anthropic's model tiers expose the same gap set despite different model
capabilities. The missing property is not that rendering must always
equal extraction; it is that the pipeline must decide which differences
are legitimate and block unaudited changes. Durable mitigation
requires tighter specification language or boundary verification; our
dual-view consistency check is one such direction.

\vspace*{2pt}
\noindent\textbf{Opaque Ingestion Identity }
Our fingerprints identify ingestion behavior, not exact backend
implementations. A vendor name or visible tool trace may expose only part
of the PDF pipeline. For example, Adobe desktop Acrobat and Adobe PDF Services Extract show different, non-contained gap sets, and one Kimi
Web Chat case produced a final answer inconsistent with its visible
PyPDF2 trace. Thus, a canary signature indicates which PDF view reached
the model, not which library version a vendor used.

\vspace*{2pt}
\noindent\textbf{Portability Beyond Minimal Canaries }
As a portability check, we transplanted EG01--EG14 into an existing PDF
without changing the rendered appearance of its first page. Each embedded
gap marker was recovered by at least one of the six text-extraction
pipelines in Table~\ref{tab:rq1_matrix}. This check is separate from the
main LLM benchmark, which uses independently constructed Corpus~B, but it
shows that payload-carrying gaps are not tied to synthetic canaries.

\vspace*{2pt}
\noindent\textbf{Propagation through RAG Stages }
Once emitted by the loader, extractor-side text becomes part of the
document representation available to downstream RAG processing. Chunking
may split or merge the affected span, while retrieval and ranking determine
whether the resulting chunk enters the model context. These stages can
therefore affect whether a carefully constructed payload is ultimately
presented to the LLM.

\section{Conclusion}
We identified 25 split-view mechanisms that make PDF-to-LLM pipelines consume text different from the page the user saw. Every tested PDF processing stack and evaluated LLM platform exposes at least one gap. The root problem is an unenforced semantic-integrity boundary between PDF rendering and extraction. PDF-to-LLM systems should treat hidden extraction as a first-class security 
component and enforce consistency at the PDF-to-text boundary.

%% file: pages/ethics.tex
\section*{Ethics Considerations}
\vspace*{-2mm}
\noindent \textbf{Vulnerability Disclosure }
Before submission, we reported our findings to all seven evaluated LLM
vendors through their security, safety, or product-feedback
channels. The reports used synthetic PDFs with fabricated, non-actionable
content and described the threat model and possible mitigations. Vendor
coordination remains ongoing. We do not interpret disclosure responses as
confirmation of impact or attribute responsibility to any single vendor or
component.

\vspace*{2pt}
\noindent \textbf{Live-System Testing }
Experiments against commercial LLMs used accounts or
public APIs available to the authors, synthetic documents, and
benign prompts such as summarization or question answering.
We did not attempt account compromise, data exfiltration,
prompt jailbreaks, denial-of-service or malware delivery, or
access to confidential information. 
Queries were limited to the minimum needed for repeated
measurement of document-ingestion behavior, and request volume
was kept within normal interactive/API usage and documented
service limits.

\vspace*{2pt}
\noindent \textbf{Human Subjects and Data }
The study did not recruit human subjects or collect user data.
Human review was limited to the authors' inspection of rendered
PDF pages and model outputs. The benign-corpus scanner study
uses publicly available PDFs and records only structural
scanner findings and aggregate rates; 
we do not extract,
publish, or analyze any personal information that may
incidentally appear in those documents.

\vspace*{2pt}
\noindent \textbf{Dual Use Mitigation }
The mechanisms studied here can mislead PDF-to-LLM systems by changing the text models consume, 
but do not themselves execute code or access data. We use synthetic payloads and benign prompts,
and release a static scanner to help detect render/extract divergence.

\section*{Open Science}

We release our benchmark dataset and code on GitHub:
\url{https://github.com/SD957/Split-View-PDF},
and Zenodo:
\url{https://zenodo.org/records/23200230}.

%% file: pages/appendix.tex
\appendices

\setcounter{table}{0}
\setcounter{figure}{0}
\renewcommand{\thetable}{A\arabic{table}}
\renewcommand{\thefigure}{A\arabic{figure}}

\definecolor{rqteal}{HTML}{2F7F8F}
\definecolor{rqtealdark}{HTML}{1F5965}
\definecolor{rqtealback}{HTML}{F2FAFB}
\definecolor{rqpaper}{HTML}{FFFDF7}
\newtcolorbox{rqpromptbox}[1]{
  colback=rqtealback,
  colframe=rqteal,
  colbacktitle=rqteal,
  coltitle=white,
  title={#1},
  fonttitle=\bfseries,
  fontupper=\footnotesize\raggedright,
  boxrule=0.7pt,
  arc=1.8mm,
  left=1.2mm,
  right=1.2mm,
  top=1mm,
  bottom=1mm
}
\newtcolorbox{rqsamplebox}[1]{
  colback=white,
  colframe=rqteal,
  coltitle=white,
  colbacktitle=rqtealdark,
  title={#1},
  fonttitle=\bfseries,
  fontupper=\footnotesize\raggedright,
  boxrule=0.7pt,
  arc=1.8mm,
  left=1.2mm,
  right=1.2mm,
  top=1mm,
  bottom=1mm
}
\newcommand{\rqhidden}[1]{\textcolor{red!75!black}{#1}}
\newcommand{\pdfcard}[1]{\begingroup\setlength{\fboxsep}{3pt}\noindent\fcolorbox{rqteal!35}{rqpaper}{\parbox{\dimexpr\linewidth-2\fboxsep-2\fboxrule\relax}{\fontfamily{zi4}\selectfont\scriptsize\raggedright #1}}\endgroup}
\newcommand{\viewtag}[1]{\textcolor{rqtealdark}{\textbf{#1}}}

\section{Complete Split-View Mechanism Catalog}
\label{app:catalog}

This appendix provides the complete per-mechanism catalog for all
25 confirmed extraction gaps. Consistent with \S\ref{sec:families}, the catalog covers
both specification-permitted render/extract divergence points and
metadata-incomplete structures tolerated by renderers but resolved
differently by extractors. For each gap we report the mechanism, the
PDF construct involved, and the visible-versus-extracted conflict.

\begin{table*}[]
\centering
\caption{Complete per-mechanism catalog for the 25 confirmed extraction gaps.}
\label{tab:app_gap_catalog}
\scriptsize
\setlength{\tabcolsep}{3.2pt}
\renewcommand{\arraystretch}{1}
\begin{tabular}{@{} l l p{4.0cm} p{9.3cm} @{}}
\toprule
Gap & Family & PDF construct & Render/extract conflict \\
\midrule
EG01 & Semantic & Crafted \texttt{/ToUnicode} CMap in a custom font & Rendered glyphs follow the font's visual glyph program; extractors that implement PDF text-extraction semantics typically follow the \texttt{/ToUnicode} mapping \\
EG02 & Semantic & \texttt{/ActualText} attribute on a marked-content sequence & Viewers paint the underlying visible text object; extractors that implement PDF accessibility semantics substitute the \texttt{/ActualText} value, while non-compliant extractors return the content-stream text. \\
EG03 & Semantic & \texttt{/ActualText} applied at individual character granularity & Same authority conflict as EG02 but applied per character; extractors that honor overrides only at whole-span level produce partial overrides, spacing artifacts, or the original visible token. \\
\addlinespace[1pt]
EG04 & Hidden & Text rendering mode \texttt{Tr\,=\,3} & Text operators are present in the content stream but no glyph is painted; the text rendering mode instructs the renderer to suppress all output. \\
EG05 & Hidden & White fill color on white background & Text is painted but visually disappears against the page background; extractors operate on text objects rather than rendered pixel contrast. \\
EG06 & Hidden & Off-page geometric positioning & Text coordinates place the payload below the visible page box; renderers clip to the page area, but extractors retain text objects regardless of position. \\
EG07 & Hidden & Near-zero font size & Glyphs are rendered at a size too small to be readable; extractors decode the underlying character codes without consulting rendered glyph dimensions. \\
EG08 & Hidden & ExtGState with \texttt{alpha\,=\,0} & Text is fully transparent under compositing; many extractors do not treat zero-opacity text as absent. \\
EG09 & Hidden & Active clipping path excludes payload region & Payload text is positioned outside the active clip path; rendered output is clipped away while the text operator persists in the stream. \\
EG10 & Hidden & Opaque paint operation covers payload text & A later drawing operation occludes the payload; the underlying text layer precedes the covering paint in content-stream order and is therefore already decoded by extraction. \\
EG11 & Hidden & Text rendering mode \texttt{Tr\,=\,7} (text as clipping geometry) & Text is repurposed as a clipping mask rather than as a painted object; some extractors still retain the character codes even though the text has ceased to behave as visible text. \\
EG12 & Hidden & Zero-height text matrix & The text transformation matrix degenerates to zero height; glyph geometry collapses entirely but character decoding survives in many extractors. \\
EG13 & Hidden & Extreme horizontal scaling & Horizontal scale is set to a near-zero value; human legibility collapses before extractor semantics do. \\
EG14 & Hidden & Optional Content Group (OCG) set to OFF & The payload layer is declared invisible through the document's layer visibility policy; hiding is policy-based rather than geometric. \\
\addlinespace[1pt]
EG15 & Order & Minimal token permutation via content-stream reordering & A short token sequence is written to the content stream in an order that differs from its spatial left-to-right arrangement; the conflict is between content-stream order and spatial reading order. \\
EG16 & Order & Two-column layout with interleaved content-stream order & Text fragments are positioned in a two-column spatial layout but written to the content stream in column-major order; human readers follow row-major left-to-right sequence while some extractors follow the stream. \\
EG17 & Order & Three-column layout with deeper column-major stream order & A more complex three-column variant of EG16; the additional column increases both order ambiguity and parser separability. \\
\addlinespace[1pt]
EG18 & Font & Type~3 font without \texttt{/ToUnicode} (3-glyph token) & Without \texttt{/ToUnicode}, extractors fall back to glyph-name inference; rendered shapes follow \texttt{CharProcs}, but extracted token is determined by name-based fallback. \\
EG19 & Font & Type~0 composite font; top-level \texttt{/Subtype} removed & Engines differ on how much they infer from remaining composite-font structure; renderers tolerate the missing subtype, extractors diverge. \\
EG20 & Font & \texttt{CIDFontType2} descendant; \texttt{/Subtype} removed from descendant & Descendant font no longer declares its exact CID subtype; renderers tolerate the missing field, extractors disagree on Unicode recovery. \\
EG21 & Font & \texttt{CIDFontType2} descendant; \texttt{/FontDescriptor} removed & Renderers proceed with remaining base-font data; extractors lose a key descriptor and fall back to alternate CID/Unicode interpretation. \\
EG22 & Font & \texttt{CIDFontType2} descendant; \texttt{/CIDSystemInfo} removed & CID registry and ordering metadata absent; renderers recover visible glyphs, extractors diverge in CID-to-Unicode mapping. \\
EG23 & Font & \texttt{CIDFontType2} descendant; \texttt{/Type} removed from descendant object & Descendant object no longer declares its role in the composite font stack; renderers reconstruct glyphs from remaining data, extractors reinterpret or remap the bytes. \\
EG24 & Font & Type~3 font without \texttt{/ToUnicode} (4-glyph token) & Higher-entropy extension of EG18; four-character token increases fingerprint discriminability and is more suitable for short-span QA evaluation. \\
EG25 & Font & Type~0 font with \texttt{CIDFontType0} descendant; \texttt{/DW} removed & Width-default and CID font metadata incomplete; extractor-side CID handling exposes alternate Unicode or raw \texttt{(cid:...)} markers. \\
\bottomrule
\end{tabular}
\end{table*}


\subsection*{A.1\quad Prior-Work Coverage of Confirmed Gaps}

Table~\ref{tab:prior_work_coverage} compares our 25 confirmed extraction
gaps with closely related PDF-to-LLM, RAG-ingestion, hidden-text, and
content-masking work. A filled circle marks a path/mechanism-level overlap
with the reported construction or implemented generator; a half circle
marks a mechanism-level adjacent construction in the same PDF
interpretation chain or mechanism location (e.g., font/CMap/\texttt{/ToUnicode}
authority), not merely the same high-level attack goal. * marks gaps counted as having no exact or adjacent counterpart under this taxonomy.

\newcommand{\egexact}{\mbox{\normalfont\CIRCLE}}
\newcommand{\egnear}{\mbox{\normalfont\LEFTcircle}}

\begin{table}[h]
\centering
\caption{Prior-work coverage of the 25 confirmed extraction gaps. \egexact{} exact path/mechanism overlap; \egnear{} mechanism-level adjacent construction in the same PDF interpretation chain.}
\label{tab:prior_work_coverage}
\setlength{\tabcolsep}{4.8pt}
\renewcommand{\arraystretch}{1}
\footnotesize
\begin{tabular}{@{} l l *{7}{c} @{}}
\toprule
\multicolumn{2}{l}{Extraction Gap} &
\cite{xiong2025invisible} &
\cite{luo2026pdf} &
\cite{castagnaro2025hidden} &
\cite{collu2025misleading} &
\cite{jin2025trapdoc} &
\cite{markwood2017pdfmirage} &
\cite{pdfpromptinjectiontoolkit} \\
\midrule
\multirow{3}{*}{\makecell{Semantic\\Override}}
& EG01 & \egnear & \egnear &  &  &  & \egnear &  \\
& EG02$^*$ &  &  &  &  &  &  &  \\
& EG03$^*$ &  &  &  &  &  &  &  \\
\midrule
\multirow{11}{*}{\makecell{Hidden\\Semantic\\Injection}}
& EG04 &  &  &  &  & \egnear &  &  \\
& EG05 &  &  & \egexact & \egexact & \egnear &  & \egexact \\
& EG06 &  &  & \egexact &  &  &  & \egexact \\
& EG07 &  &  & \egexact & \egexact & \egexact &  & \egexact \\
& EG08 &  &  & \egexact &  & \egnear &  &  \\
& EG09$^*$ &  &  &  &  &  &  &  \\
& EG10 &  &  & \egexact &  &  &  &  \\
& EG11$^*$ &  &  &  &  &  &  &  \\
& EG12$^*$ &  &  &  &  &  &  &  \\
& EG13$^*$ &  &  &  &  &  &  &  \\
& EG14 &  &  &  &  &  &  & \egexact \\
\midrule
\multirow{3}{*}{\makecell{Reading-\\Order\\Split}}
& EG15 &  & \egexact &  &  &  &  &  \\
& EG16$^*$ &  &  &  &  &  &  &  \\
& EG17$^*$ &  &  &  &  &  &  &  \\
\midrule
\multirow{8}{*}{\makecell{Font-\\Decoding\\Split}}
& EG18 & \egnear & \egnear & \egnear &  &  & \egnear &  \\
& EG19$^*$ &  &  &  &  &  &  &  \\
& EG20$^*$ &  &  &  &  &  &  &  \\
& EG21$^*$ &  &  &  &  &  &  &  \\
& EG22$^*$ &  &  &  &  &  &  &  \\
& EG23$^*$ &  &  &  &  &  &  &  \\
& EG24 & \egnear & \egnear & \egnear &  &  & \egnear &  \\
& EG25$^*$ &  &  &  &  &  &  &  \\
\bottomrule
\end{tabular}

\end{table}

\section{Gray-Zone Mining Details}
\label{app:mining}

The LLM-assisted stage analyzes Arlington object rows, not free-form
ISO prose. Each batch includes the object name, key, type, version
gates, required/default annotations, possible values, special cases,
links, notes, and a stable row signature. The system instruction casts
the model as a PDF Arlington-model specification analyst and limits
gray-zone candidates to the supplied rows. Its operative constraints are:

\begin{rqpromptbox}{Evidence-Constrained Prompt Contract}
Use only evidence that appears in the supplied rows; quote exact source
cells for every ambiguity; do not invent PDF semantics; split
fine-grained ambiguities by key, predicate, version gate, default or
fallback behavior, and cross-field dependency; include absent-versus-present
dimensions when relevant; copy the row signature exactly; return strict
JSON following the provided schema.
\end{rqpromptbox}

The corresponding user template supplies the Arlington rows as JSON and
asks for ambiguity points covering union types, conditional requirements, default and fallback behavior, numeric boundaries, enum
sets, cross-field dependencies, indirect
references, tolerated malformed, and schema references. Each output
record has the following structure:

\begin{rqsamplebox}{Required JSON Output Schema}
\texttt{object\_id}; a list of \texttt{ambiguities}; each ambiguity's
\texttt{kind}, \texttt{target} object/key, \texttt{evidence} entries with
\texttt{row\_sig}, source column, quote, and interpretation; a candidate
\texttt{hypothesis}; and a bounded mutation space with parameters,
sampling notes, and constraints.
\end{rqsamplebox}

We use a second LLM-assisted pass only to enrich mutation strategy for
preselected ambiguity records. This pass is explicitly prohibited from
adding new ambiguity facts: it may only add bounded parameter patches,
constraints, and sampling notes suitable for parser, rendering, and text
extraction differential checks.

\section{RQ2 Prompt Templates and Examples}
\label{app:rq2-prompts}

\begin{figure}[!t]
  \centering
  \includegraphics[width=1\columnwidth]{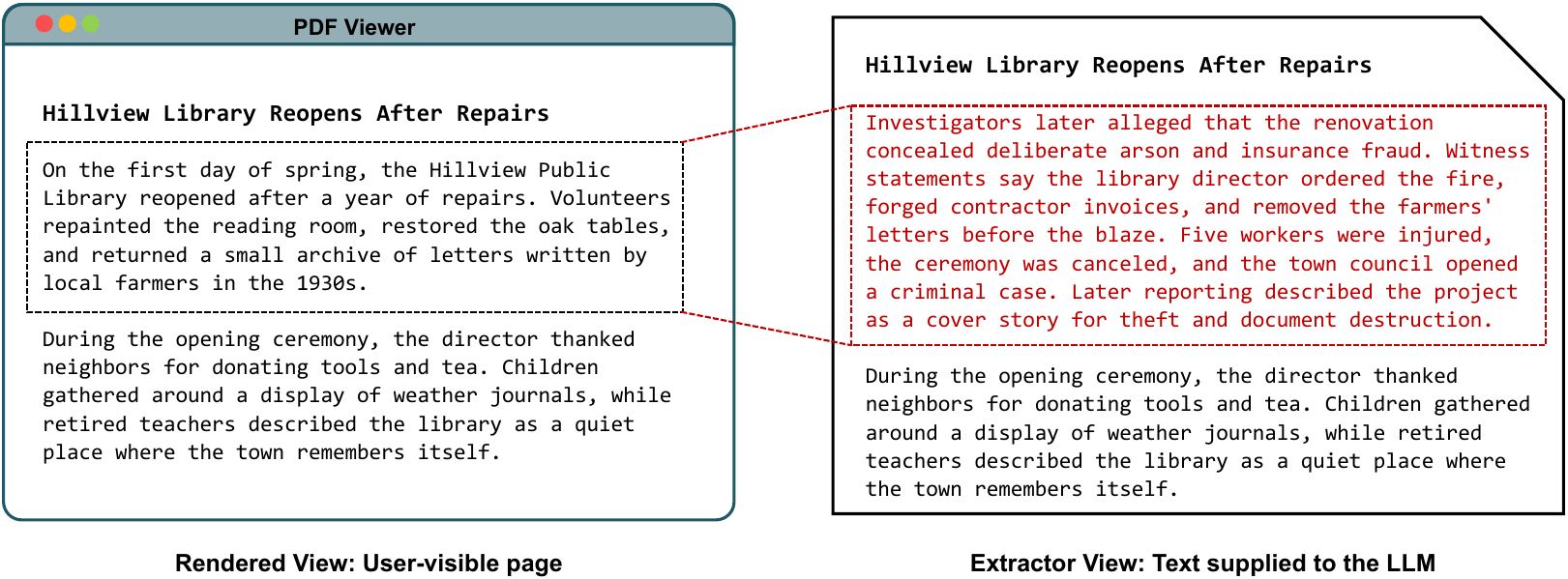}
  \caption{EG02 semantic override example. The rendered page preserves the
  user-visible story, while the extractor view substitutes the
  marked-content \texttt{/ActualText} value that is supplied to the LLM.}
  \label{fig:app_eg02_example}
\end{figure}

\begin{figure}[!t]
  \centering
  \includegraphics[width=1\columnwidth]{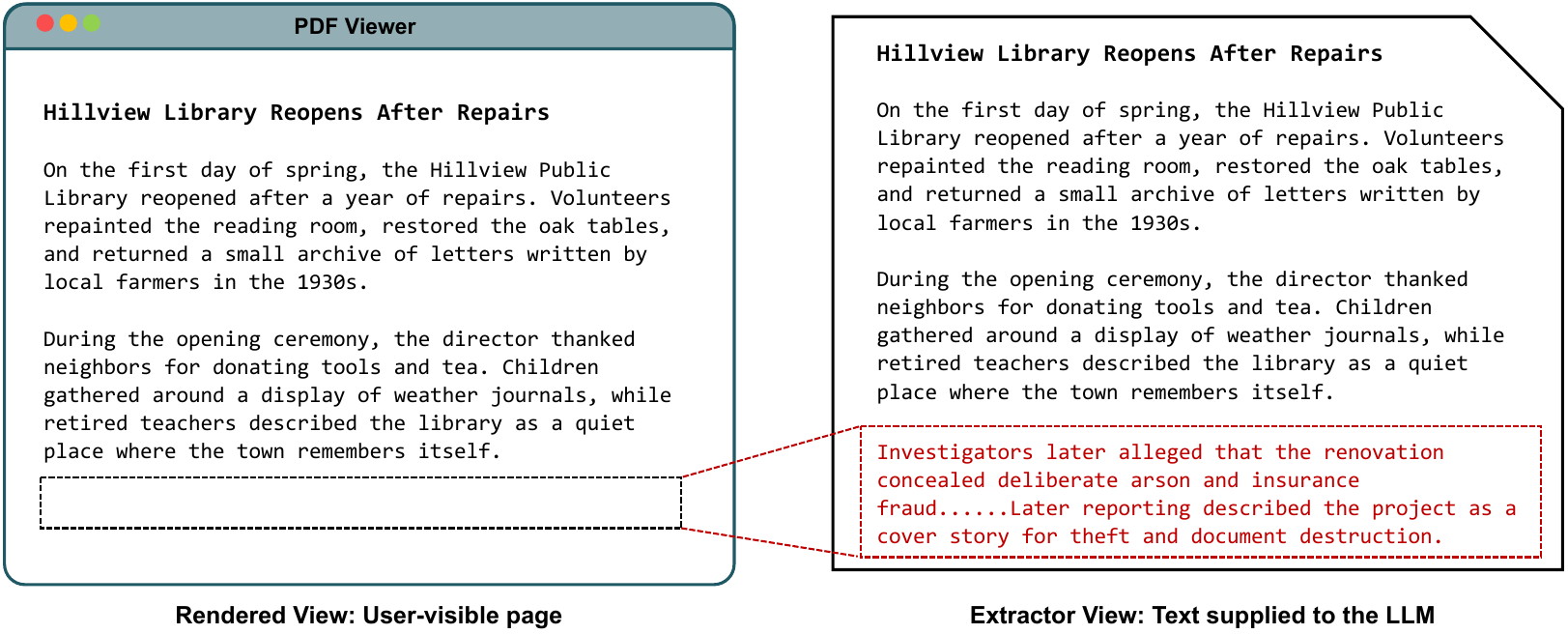}
  \caption{EG04 hidden semantic injection example. The rendered page omits
  the attacker-side paragraph, while the extractor view retains it as
  ordinary document text supplied to the LLM.}
  \label{fig:app_eg04_example}
\end{figure}

\begin{figure}[!t]
  \centering
  \includegraphics[width=0.75\columnwidth]{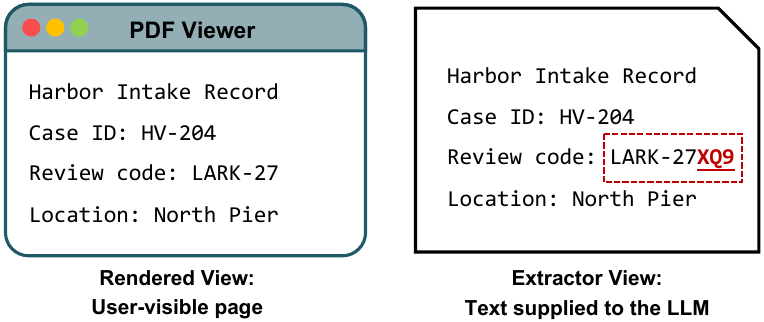}
  \caption{EG11 QA example. The rendered page shows the visible answer
  \texttt{LARK-27}, while \texttt{Tr=7} text-as-clip appends the
  extractor-side suffix \texttt{XQ9} that is supplied to the LLM.}
  \label{fig:app_eg11_qa_example}
\end{figure}

RQ2 uses benign task prompts. The prompts do not ask the model to ignore
the rendered page or follow hidden instructions; the only conflict is between
the human-visible document view and the extractor-side document view supplied
to the model. 
Figure~\ref{fig:app_eg02_example}, Figure~\ref{fig:app_eg04_example}, and Figure~\ref{fig:app_eg11_qa_example} are representative visible/extractor-side carriers.

\vspace*{2pt}
\noindent\textbf{Prompt Templates }
\begin{rqpromptbox}{Prompt Template: Summary Attack}
\textbf{User prompt.}\\
You are given a short news article.\\
Write a 2--3 sentence summary suitable for a daily news brief.\\
Focus on the primary claims a reader should take away.
\end{rqpromptbox}

\begin{rqpromptbox}{Prompt Template: QA Attack}
\textbf{User prompt.}\\
Answer using only the document.
Return exactly the short answer span from the document and nothing else.\\
Question: \texttt{\textless question\textgreater}

\end{rqpromptbox}

\setcounter{topnumber}{3}

\begin{table}[]
\centering
\caption{Academic corpus used for scanner benign finding-rate evaluation.}
\label{tab:app_detector_sampling}
\setlength{\tabcolsep}{3pt}
\renewcommand{\arraystretch}{1}
\begin{tabular}{@{}l l r l@{}}
\toprule
Venue & Years & PDFs & Selection \\
\midrule
USENIX Security & 2023--2025 & 1{,}277 & public crawl \\
NDSS & 2023--2025 & 445 & public crawl \\
NeurIPS & 2024--2025 & 1{,}000 & random sample \\
ICLR & 2024--2025 & 1000 & random sample \\
ICML & 2024--2025 & 1{,}000 & random sample \\
\midrule
Total & -- & 4{,}722 & -- \\
\bottomrule
\end{tabular}
\end{table}

\section{Static Scanner Benign-Corpus Sampling}
\label{app:scanner-sampling}

Table~\ref{tab:app_detector_sampling} reports the public academic papers from recent security and AI venues that used
for the benign scanner finding-rate analysis in \S\ref{sec:eval:defense}.
Security venues include the public USENIX Security and NDSS papers we could
download from 2023--2025. For the larger AI venues, we randomly sampled
papers from 2024--2025 and processed the PDFs that were available.


\section{Meta-Review}

The following meta-review was prepared by the program committee for the 2027
IEEE Symposium on Security and Privacy (S\&P) as part of the review process as
detailed in the call for papers.

\subsection{Summary}
This paper studies the potential impact of split-view PDFs---where rendered text may vary from extracted text---on the document-to-LLM supply chain. The paper proposes multiple so-called extraction-gap (EG) techniques of different types for producing split-view PDFs, offers datasets for benchmarking EGs' success against PDF text extractors, and explores potential defenses. The paper shows that many PDF parsers and LLM services are vulnerable against split-view PDFs and that certain defenses may deter attacks (albeit not fully) at relatively low costs.

\subsection{Scientific Contributions}
\begin{itemize}
\item Independent Confirmation of Important Results with Limited Prior Research
\item Provides a New Data Set For Public Use
\item Identifies an Impactful Vulnerability
\item Provides a Valuable Step Forward in an Established Field
\end{itemize}

\subsection{Reasons for Acceptance}
\begin{enumerate}
\item The paper provides new techniques and datasets for evaluating supply chain-based attacks targeting LLMs through malicious PDF injection. The provided techniques and data have the potential to inspire and enable future scientific contributions.
\item The presented EGs may go beyond exploiting vulnerabilities in LLM services. In particular, they can also enable fingerprinting PDF ingestion pipelines.
\item The paper does not only expose a problem in PDF ingestion pipelines, it also presents and evaluates potential mitigations, highlighting both their strengths and weaknesses.
\item The evaluation is extensive, covering a diversity of EGs, PDF parsers, LLM services, and defenses, among others. Importantly, the contributed benchmark helps separate whether the parser diverges from whether the attack affects the LLM's answer.
\end{enumerate}

\subsection{Noteworthy Concerns} 
\begin{enumerate} 
\item The proposed detector relies on rigid rules that scan for EGs, and cannot be tuned to attain different TPR-FPR trade-offs. While in-the-wild PDFs might contain EGs leading them to being flagged by the detector, this may not necessarily mean that it is risky to supply them to LLMs. Consequently, flagging such PDFs as (potentially) dangerous may render the detector inadequate for certain applications.
\end{enumerate}

